
\newif\iflanl
\openin 1 lanlmac
\ifeof 1 \lanlfalse \else \lanltrue \fi
\closein 1
\iflanl
    \input lanlmac
\else
    \message{[lanlmac not found - use harvmac instead}
    \input harvmac
\fi
\newif\ifhypertex
\ifx\hyperdef\UnDeFiNeD
    \hypertexfalse
    \message{[HYPERTEX MODE OFF}
    \def\hyperref#1#2#3#4{#4}
    \def\hyperdef#1#2#3#4{#4}
    \def\hypernoname{}
    \def\e@tf@ur#1{}
    \def\hep-th/#1#2#3#4#5#6#7{{\tt hep-th/#1#2#3#4#5#6#7}}
\else
    \hypertextrue
    \message{[HYPERTEX MODE ON}
  \def\hep-th/#1#2#3#4#5#6#7{
  {\tt hep-th/#1#2#3#4#5#6#7}}
\fi
\newif\ifdraft

\noblackbox
\catcode`\@=11
\newif\iffrontpage
\newif\iffigureexists
\newif\ifepsfloaded
\def\epsfcheck{
\ifdraft
\input epsf\epsfloadedtrue
\else
  \openin 1 epsf
  \ifeof 1 \epsfloadedfalse \else \epsfloadedtrue \fi
  \closein 1
  \ifepsfloaded
    \input epsf
  \else
\immediate\write20{NO EPSF FILE --- FIGURES WILL BE IGNORED}
  \fi
\fi
\def\epsfcheck{}}
\def\checkex#1{
\ifdraft
\figureexistsfalse\immediate%
\write20{Draftmode: figure #1 not included}
\else\relax
    \ifepsfloaded \openin 1 #1
	\ifeof 1
           \figureexistsfalse
  \immediate\write20{FIGURE FILE #1 NOT FOUND}
	\else \figureexiststrue
	\fi \closein 1
    \else \figureexistsfalse
    \fi
\fi}
\def\missbox#1#2{$\vcenter{\hrule
\hbox{\vrule height#1\kern1.truein
\raise.5truein\hbox{#2} \kern1.truein \vrule} \hrule}$}
\def\lfig#1{
\let\labelflag=#1%
\def\numb@rone{#1}%
\ifx\labelflag\UnDeFiNeD%
{\xdef#1{\the\figno}%
\writedef{#1\leftbracket{\the\figno}}%
\global\advance\figno by1%
}\fi{\hyperref{}{figure}{{\numb@rone}}{Fig.{\numb@rone}}}}
\def\figinsert#1#2#3#4{
\epsfcheck\checkex{#4}%
\def\figsize{#3}%
\let\flag=#1\ifx\flag\UnDeFiNeD
{\xdef#1{\the\figno}%
\writedef{#1\leftbracket{\the\figno}}%
\global\advance\figno by1%
}\fi
\goodbreak\midinsert%
\iffigureexists
\centerline{\epsfysize\figsize\epsfbox{#4}}%
\else%
\vskip.05truein
  \ifepsfloaded
  \ifdraft
  \centerline{\missbox\figsize{Draftmode: #4 not included}}%
  \else
  \centerline{\missbox\figsize{#4 not found}}
  \fi
  \else
  \centerline{\missbox\figsize{epsf.tex not found}}
  \fi
\vskip.05truein
\fi%
{\smallskip%
\leftskip 4pc \rightskip 4pc%
\noindent\ninepoint\sl \baselineskip=11pt%
{\bf{\hyperdef\hypernoname{figure}{{#1}}{Fig.{#1}}}:~}#2%
\smallskip}\bigskip\endinsert%
}
%
%

\def\LG{Lan\-dau-Ginz\-burg\ }
\def\ie{{\it i.e.} }

\def\half{{1 \over 2}}

\def\sk3{{\sqrt{k+3}}}
\def\half{{1 \over 2}}

\def\ssk2{{\sqrt{k+2}}}
\def\s3{{\sqrt{3}}}

\def\advp#1{{\it Adv.\ Phys.}\ {\bf #1\/}}
\def\annp#1{{\it Ann.\ Phys.}\ {\bf #1\/}}
\def\cmp#1{{\it Commun.\ Math.\ Phys.} \ {\bf #1\/}}
\def\nup#1{{\it Nucl.\ Phys.} \ {\bf B#1\/}}
\def\plt#1{{\it Phys.\ Lett.}\ {\bf #1\/}}
\def\ijmp#1{{\it Int.\ J.\ Mod.\ Phys.}\ {\bf A#1\/}}
\def\jpa#1{{\it J.\ Phys.}\ {\bf A#1\/}}

\def\prl#1{{\it Phys.\ Rev.\ Lett. }\ {\bf #1}\/}
\def\prb#1{{\it Phys.\ Rev.} \ {\bf B#1}\/}
\def\prd#1{{\it Phys.\ Rev.}\ {\bf D#1}\/}

\def\mpl#1{{\it Mod.\ Phys.\ Lett.}\   {\bf A#1}\ }
\def\rmp#1{{\it Rev.\ Mod.\ Phys.}\ {\bf #1}\/}
\def\jetp#1{{\it Sov.\ Phys.\  JETP} \ {\bf #1}\/}

\def\Gminus{G_{-{1 \over 2}}^-}
\def\Gplus{G_{-{1 \over 2}}^+}

\def\Gmp{G_{-{1 \over 2}}^\mp}

\def\coeff#1#2{\relax{\textstyle {#1 \over #2}}\displaystyle} 
\def\inbar{\vrule height1.5ex width.4pt depth0pt}
\def\IC{\relax\,\hbox{$\inbar\kern-.3em{\rm C}$}}
\def\IR{\relax{\rm I\kern-.18em R}}
\font\sanse=cmss12
\def\ZZ{\relax{\hbox{\sanse Z\kern-.42em Z}}}

\def\shalf{\coeff{1}{2}}
\noblackbox
\def\MyTitletwo#1{\nopagenumbers\abstractfont\hsize=\hstitle
\rightline{}
\vskip .7in
\centerline{\titlefont #1}
\abstractfont\vskip .5in\pageno=0}

\def\Date#1{\leftline{#1}\tenpoint\supereject\global\hsize=\hsbody%
\footline={\hss\tenrm\folio\hss}}
%

%
\lref\GZam{S.~Ghoshal and A.B.~Zamolodchikov, \ijmp{9}
(1994) 3841,  \hep-th/9306002.}
\lref\Kondo{N.~Andrei, K~Furuya and J.~Lowenstein, \rmp{55}
(1983) 331; \hfill \break
A.M.~Tsvelik and P.B.~Wiegmann, \advp{32} (1983) 453.}
\lref\durham{E.~Corrigan, P.E.~Dorey,  R.H.~Rietdijk and R.~Sasaki,
``Affine Toda Field Theory on a Half--Line,'' \plt{333B} (1994) 83;
DTP-94-57, \hep-th/9501098;  \hfill \break
P.~Bowcock, E.~Corrigan, P.E.~Dorey and R.H.~Rietdijk, ``Classically
Integrable Boundary Conditions for Affine Toda Field Theories,''
DTP-94-57, \hep-th/9501098.}
\lref\itals{S.~Penati and D.~Zanon  ``Quantum Integrability in
Two--Dimensional Systems with Boundary,'' IFUM-490-FT,
\hep-th/9501105.}
\lref\CalRub{V.A.~Rubakov, \nup{203} (1982) 311; \hfill \break
C.G.~Callan,   \prd{25} (1982)  2141; \prd{26} (1982) 2058;
\nup{212} (1983) 391;\hfill \break  C.G.~Callan and S.R.~Das,
\prl{51} (1983) 1155 .}
\lref\Wen{X.G.~Wen, \prb{41} (1990) 12838;
\prb{43} (1991) 11025; \prb{44} (1991) 5708. }
\lref\FLS{P.~Fendley, A.~Ludwig and H.~Saleur, \prl{74} (1995) 3005;
cond-mat/9408068.}
\lref\OW{E.~Witten and D.~Olive, \plt{78} (1978) 97.}
\lref\FMVW{P.~Fendley, S.~Mathur, C.~Vafa and N.P.~Warner, \plt{243B}
(1990) 257.}
\lref\WLNW{W.~Lerche and N.P.~Warner, \nup{358} (1991) 571.}
\lref\EllGen{E.~Witten, \ijmp{9} (1994) 4783; \hfill \break
P.~di Francesco and S.~Yankielowicz, \nup{409} (1993) 18;
\hfill \break P.~di Francesco, O.~Aharony and S.~Yankielowicz,
\nup{411} (1994) 584; \hfill \break
T.~Kawai, Y.~Yamada and S.-K.~Yang, \nup{414} (1994) 191;
\hfill \break  T.~Kawai, \plt{342B} (1995) 87; T.~ Kawai and
S.-K.~ Yang, ``Duality of Orbifoldized Elliptic
Genera,'' preprint KEK-TH-409, \hep-th/9408121; \hfill \break
M.~Henningson, \nup{413} (1994) 73;
\hfill \break  P.~Berglund and M.~Henningson,
\nup{433} (1995) 311; ``On the Elliptic Genus and Mirror Symmetry,''
IASSNS-HEP-94-11, \hep-th/9406045.}
\lref\EGenNW{D.~Nemeschansky and N.P.~Warner, \plt{329B} (1994) 53;
``The Refined Elliptic Genus and Coulomb Gas Formulations of $N=2$
Superconformal Coset Models,''  USC-94-018, \hep-th/9412187.}
\lref\GenInd{S.~Cecotti, P.~Fendley, K.~Intriligator and C.~Vafa,
\nup{386} (1992) 405; \hfill \break
P.~Fendley and H.~Saleur, \nup{388} (1992) 609.}
\lref\Cardy{J.~Cardy, \nup{240} (1984) 514; \nup{275} (1986) 200;
\nup{324},(1989) 581.}
\lref\Cardy{J.L.~Cardy, \nup{240} (1984) 514; \nup{275} (1986) 200;
\nup{324},(1989) 581; \hfill \break
J.L.~Cardy and D.C.~Lewellen, \plt{259B} (1991) 274.}
\lref\Zamo{A.B. Zamolodchikov, {\it JETP Letters} {\bf 46}
(1987) 161; ``Integrable field theory from conformal field theory''
in {\it Proceedings of the Taniguchi symposium} (Kyoto 1989) and in
{\it Adv. Studies in Pure Math.} 19 (1989) 641; R.A.L. preprint
89-001; \ijmp{4} (1989) 4235.}
\lref\Toda{A.~Bilal and J.-L.~Gervais, \plt{206B} (1988) 412;
\nup{314} (1989) 646; \nup{318} (1989) 579; \nup{326} (1989) 222.
\hfill \break  A.\ Bilal, \nup{330} (1990) 399;  \ijmp{5} (1990)
1881. \hfill \break
T.~Hollowood and P.~Mansfield, \plt{226B} (1989) 73; \nup{330}
(1990) 720. \hfill \break
T.~Eguchi and S.-K.~Yang, \plt{224B} (1989) 373; \plt{235B} (1990)
282.}
\lref\ABL{D. Bernard and A. LeClair, \plt{247B} (1990) 309; \nup{340}
(1990) 721; \cmp{142} (1991) 99; \hfill \break
C. Ahn, D. Bernard, and A. LeClair, \nup{346} (1990) 409. }
\lref\FLMW{P.~Fendley, W.~Lerche, S.D.~Mathur and N.P.~Warner,
\nup{348}
(1991) 66.}
\lref\susyint{ P.~Mathieu and M.A.~Walton, \plt{254B} (1991) 106;
\hfill \break
M.T.~Grisaru, S.~Penati and D.~Zanon, \plt{253B} (1991) 357; \hfill
\break
G.W.~Delius, M.T.~Grisaru, S.~Penati and D.~Zanon, \plt{256B}
(1991) 164; \nup{359} (1991) 125; \hfill \break
J.~Evans and T.J.~Hollowood \nup{352} (1991) 723;
\nup{382} (1992) 662; \plt{293B} (1992) 100.}
\lref\CGints{V.S.~Dotsenko and V.A.~Fateev, \plt{154B}(1985) 291;
\nup{251} (1985) 691; \hfill \break J.~Bagger, D.~Nemeschansky and
J.-B.~Zuber, \plt{216B} (1989) 320; \hfill\break S.D.~Mathur,
\nup{369} (1992) 433.}
\lref\ntwoLG{E.~Martinec, \plt{217B} (1989) 431; \hfill\break
C.~Vafa and N.P.~Warner, \plt{218B} (1989) 51; \hfill\break
E.~Martinec, {\it ``Criticality, catastrophes and
compactifications,''}  V.G. Knizhnik memorial volume, L.~Brink {\it
et al.} (editors): {\it Physics and mathematics of strings.} }
\lref\NWTrieste{N.P. Warner,
``$N=2$ Supersymmetric Integrable Models and Topological Field
Theories,''  in Trieste 1992, Proceedings, {\it High Energy Physics
and Cosmology,} pp 143-179.  Edited by E.~Gava, K.~Narain,
S.~Randjbar-Daemi, E.~Sezgin, and  Q.~Shafi, World Scientific, 1993;
ICTP Series in Theoretical Physics, v. 9. }
\lref\PFKI{P.~Fendley and K.~Intriligator, \nup{372} (1992) 533;
\nup{380} (1992) 265.}
\lref\LNW{A.~LeClair, D.~Nemeschansky and N.P.~Warner, \nup{390}
(1993) 653.}
\lref\ZandZ{A.B.~Zamolodchikov and Al.B.~Zamolodchikov, \annp{120}
(1979) 253.}
\lref\SSW{H.~Saleur, S.~Skorik and N.P.~Warner, ``The Boundary
Sine--Gordon Theory: Classical and Semi-Classical Analysis,''
U.S.C. preprint USC-94-013, \hep-th/9408004.}
\lref\FSW{P.~Fendley, H.~Saleur and N.P.~Warner,
\nup{430} (1994) 577.}
\lref\Wita{E.~Witten, \nup{185} (1981) 513; \nup{202} (1982) 253.}
\lref\Ghosh{S.~Ghoshal, \ijmp{9} (1994) 4801; hep-th/9310188.}
\lref\HLuck{H.~Luckock, \jpa{24} (1991) L1057-L1064. }
\lref\topLG{C.~Vafa, \mpl{6}  (1991) 337.}
\lref\PWAF{P.B.~Wiegmann and A.M.~Finkelstein, \jetp{75} (1978)
204; {\bf 48} (1978) 1.}
\lref\PFKondo{P.~Fendley, \prl{71} (1993) 2485, cond-mat/9304031.}
\lref\FGLS{P.~ Fr\'e, L.~Girardello,  A.~ Lerda and P.~Soriani,
\nup{387} (1992) 333.}
\lref\Witb{E.~Witten, \ijmp{9} (1994) 4783.}
%
%
%
%
%


\MyTitletwo
{Supersymmetry in Boundary Integrable Models
$^*$ { \abstractfont
\footnote{}{$^*$ Work supported in part by funds provided by the DOE
under grant No. DE-FG03-84ER40168.}} }  {}
\centerline{N.P. Warner}
\bigskip
\centerline{Physics Department, University of Southern California}
\centerline{University Park,  Los Angeles, CA 90089--0484.}
\medskip
\centerline{and}
\medskip
\centerline{{\it LPTHE\/,} \
Universit\'e Pierre et Marie Curie - Paris VI}
\centerline{Universit\'e Denis Diderot - Paris VII}
\centerline{Laboratoire associ\'e No. 280 au CNRS $^\dagger$
{\abstractfont \footnote{}{ $^\dagger$  Boite 126, Tour 16,
1$^{er}$ \'etage, 4 place Jussieu, F-75252 Paris CEDEX 05,
FRANCE.}}}

\vskip 1.0 cm

Quantum integrable models that possess $N=2$ supersymmetry are
investigated on the half-space. Conformal perturbation theory is used
to identify some $N=2$ supersymmetric boundary integrable models, and
the effective boundary Landau-Ginzburg formulations are constructed.
It is found that $N=2$ supersymmetry largely determines the boundary
action in terms of the bulk, and in particular, the boundary bosonic
potential is $|W|^2$, where $W$ is the bulk superpotential.
Supersymmetry is also investigated using the affine quantum group
symmetry of exact scattering matrices, and the affine quantum group
symmetry of boundary reflection matrices is analyzed both for
supersymmetric and more general models. Some $N=2$
supersymmetry preserving boundary reflection matrices are given, and
their connection with the boundary Landau-Ginzburg actions is
discussed.

\vfill
\leftline{USC-95/014\/, \  hep-th/9506064}
\Date{June 1995}
%


\newsec{Introduction}

The study of conformal and quantum integrable models in
two-dimensional systems with a boundary is not only an intrinsically
interesting subject, but also provides yet another avenue by which
one can relate integrable models to physically relevant systems and,
in particular to situations that are of interest in three or four
dimensional field theory. The basic two-dimensional boundary model
consists of a system defined upon the half-line $(x<0)$ with a
spatial boundary at $x=0$. The field theory in the bulk ({\it i.e.}
for $x<0$) can be either conformal or quantum integrable, while the
boundary conditions can be simple (free or fixed), or non-trivial,
involving a mass scale and possibly non-linear boundary couplings and
dynamics. For the boundary model to be integrable (\ie possess higher
spin conserved charges), there are stringent constraints upon the
bulk and boundary sectors (for example, see \refs{\GZam, \Kondo,
\durham,\itals}).

Probably the most famous physical example of a boundary integrable
model occurs in the Kondo problem, where the $(1+1)$-dimensional
field theory is the effective field theory of $s$-wave scattering of
electrons off a magnetic spin impurity. The bulk theory is massless
(conformal) while the boundary has a mass scale (the Kondo
temperature).  Such impurity problems, in which one concentrates on
$s$-wave scattering off some isolated object at the origin,
generically provide interesting $(1+1)$-dimensional boundary field
theories. Some of these may be quantum integrable, or may have
physically useful approximations that are integrable. Another example
of this is the Callan-Rubakov effect \CalRub, where scattering of
fermions off a monopole is reduced to the study of free fermions on
the half-line.

There are also intrinsically $(1+1)$-dimensional
boundary quantum integrable systems of experimental relevance: There
is a significant body of evidence to support a Luttinger liquid
(Thirring) model for the edge states of electrons in the fractional
quantum Hall effect \Wen, and the boundary sine-Gordon model thus
provides an accurate description of the conductance though a point
contact \FLS.

In this paper I will consider supersymmetry in boundary integrable
models.  The idea is to start with an $N=2$ superconformal model on
the plane, and to see to what extent supersymmetry can be preserved
once one passes to the half-space and perturbs with relevant
boundary, and possibly bulk, operators that lead to a quantum
integrable theory.  There are several motivations for considering
this issue.  First, if one has $N=2$ supersymmetry on the plane or
cylinder, then one can obtain quite a number of exact quantum
properties of such models via semi-classical analysis:  One can
determine the vacuum structure and soliton mass ratios from an
effective  \LG formulation \refs{\OW, \FMVW, \WLNW}; one can
compute some partition functions from the elliptic genus
\refs{\EllGen,\EGenNW}, and get some exact scaling functions from the
generalized index \GenInd.  It is thus natural to investigate whether
one will be able to generalize these sorts of results
to the half-space.  At a more formal level, the supersymmetric
models, and their topological counterparts, provide the simplest
examples of Coulomb gas methods, along with the associated action of
the affine quantum group on the soliton spectrum and $S$-matrices.
Once again, one would like to know how much of these structures
survive for quantum integrable models on the half-space.  Finally,
from the point of view of higher dimensional field theories, if one
considers monopoles or strings in a supersymmetric field
theory, and treats them as impurity problems, then the result will be
supersymmetric $(1+1)$-dimensional boundary field theories.  In
some limit these will be superconformal models, and if one is
lucky, in some  approximation where mass scales are re-introduced,
one might get an integrable field theory.  It is thus important
to study supersymmetry in boundary field theories, and
investigate to what extent the supersymmetry is broken by the
presence of the boundary, and which of these models are quantum
integrable.

I will start in section 2 by reviewing some pertinent facts about
conformal field theory and extended chiral algebras on the
half-space. In section 3, I review the conditions under which
boundary and bulk perturbations lead to quantum integrable models,
and section 4 contains an analysis of the $N=2$ supersymmetric
boundary integrable models that can be obtained from perturbations of
the $N=2$ superconformal minimal series. In particular, it is shown
how families of such models can be obtained from the bulk
perturbations that lead to superpotentials that are Chebyshev
polynomials. The second major thread of this work is introduced in
section 5, where exact scattering matrices and their affine quantum
group symmetries are employed in an examination of the supersymmetry
of one of the simplest supersymmetric boundary integrable models. In
section 6, I construct some exact boundary reflection matrices that
have both free parameters and preserve $N=2$ supersymmetry. Finally,
section 7 gives the construction of the effective \LG theories for
the models described in the earlier sections, and in particular it is
found that if the bulk superpotential is $W(\phi)$, then the boundary
superpotential, $V$, satifies ${\partial V \over \partial \phi} = W$,
and hence the boundary bosonic potential is $|W|^2$.

\newsec{Conformal Systems with a Boundary}

Consider a conformal field theory on the complete complex plane, and
suppose that the theory is symmetric between the left-moving and
right-moving sectors.  In particular, this means that every
left-moving operator, ${\cal O}(z)$, in the chiral algebra ${\cal
A}$, has a right-moving counterpart, $\widetilde {\cal O}(\bar z)$,
in the chiral algebra $\widetilde {\cal A} \equiv {\cal A}$.  As is
familiar in open string theory, the introduction of a boundary
requires that the left-movers and right-movers be locked together.
As a result, the two chiral algebras,   ${\cal A}$ and $\widetilde
{\cal A}$, become identified, producing a single copy,  $\hat
{\cal A}$, of  ${\cal A}$ in the system with the boundary.  Perhaps
the simplest way to think of this is as a generalized method of
images:  That is, ${\cal A}$-preserving boundary conditions require
\Cardy:
\eqn\Abcond{{\cal O}(z) \big |_{x=0}  ~=~ \widetilde {\cal O}(- \bar
z) \big |_{x=0} \ .}
One can then think of $\widetilde {\cal O}(- \bar z)$ as the analytic
continuation of ${\cal O}(z)$ into $x = {\cal R}e (z) > 0$.

To define the conserved charges, ${\cal O}_n$, it is convenient to
introduce semi-circular contours $C_L$ and $C_R$, centered on the
origin, $O$. These contours are mirror images of each other through
the $y$-axis, and have bases that run along the $y$-axis (see
\lfig\figone). Let $C$ be the complete circular contour about $O$,
made up of the semi-circular arcs of $C_L$ and $C_R$. Then define:
$$
\eqalign{{\cal O}_n   ~=~ & \coeff{1}{2 \pi i}~ \oint_{C_L} ~
z^{n+p-1} {\cal O}(z) ~dz ~-~  \coeff{1}{2 \pi i}~ \oint_{C_R} ~
(-\bar z)^{n+p-1} \widetilde {\cal O}(- \bar z)  ~d(- \bar z) \cr
{}~=~ & \coeff{1}{2 \pi i}~ \oint_{C} ~ z^{n+p-1} {\cal O}(z) ~dz \
,}
$$
where $p$ is the conformal weight of ${\cal O}(z)$.

\figinsert\figone{The contours for defining modes of operators
on the half space.}{2.5in}{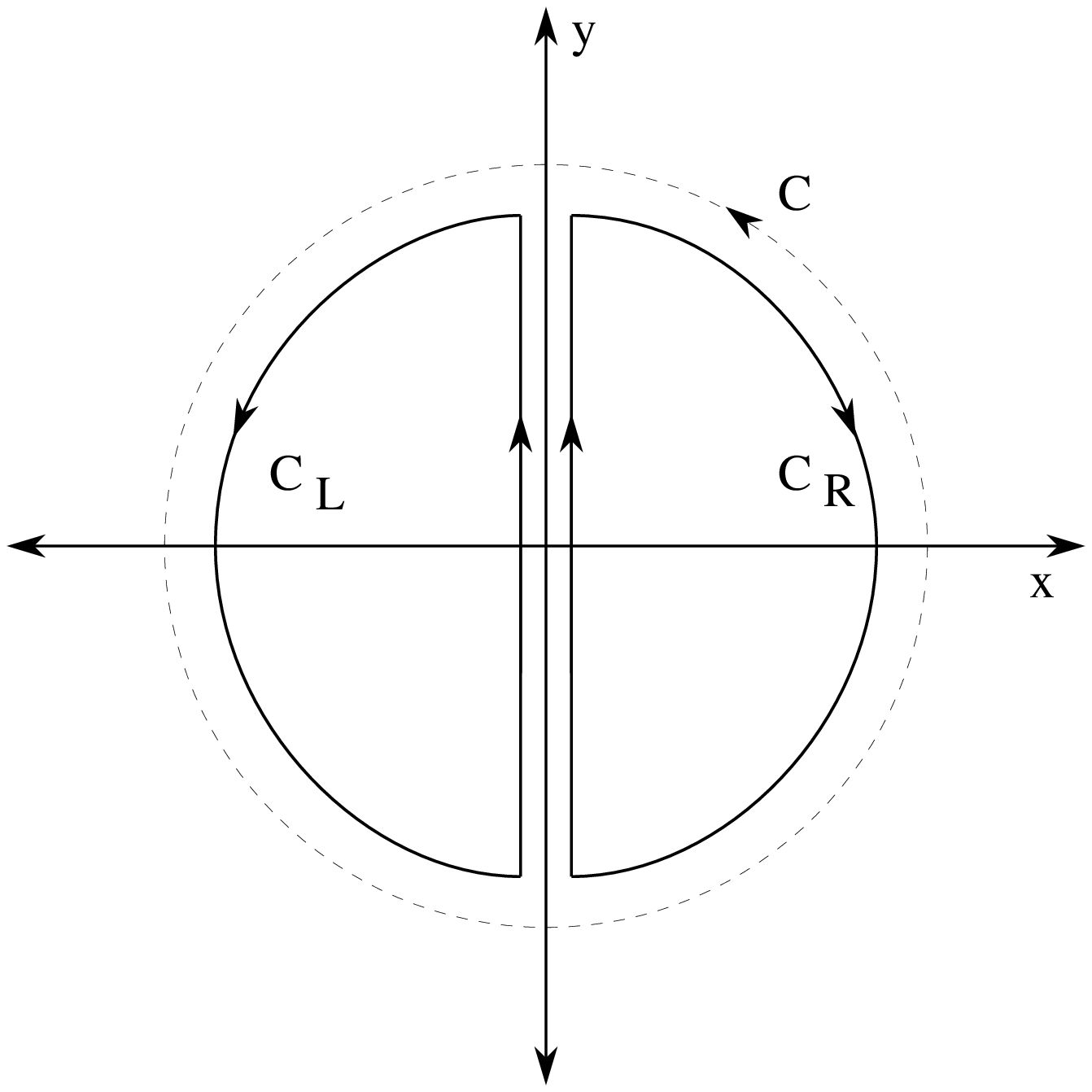}

For $N=2$ superconformal boundary
conditions one requires:
$$
\eqalign{ J (z) \big |_{x=0}  ~&=~  \ \widetilde J (- \bar z)
\big |_{x=0} \ , \cr  T (z) \big |_{x=0}  ~&=~  \ \widetilde
T (- \bar z) \big |_{x=0} \ , \cr  G^\pm (z) \big |_{x=0} ~&=~
\ \widetilde G^\pm (- \bar z) \big |_{x=0} \ .}
$$
The result is a single $N=2$ superconformal algebra on the
half-space. The choice of pairing $G^\pm (z)$ with $\widetilde G^\pm
(- \bar z)$ or with $\widetilde G^\mp (- \bar z)$ is a matter of
convention, but changing this pairing will introduce a relative
negative sign in the pairing of $J(z)$ and $\widetilde J(z)$.

One should also note that the
boundary conditions \Abcond\ do not completely determine the
boundary conditions for the individual primary fields in the theory,
but merely guarantee that the theory on the half space is ${\cal
A}$-conformal (\ie conformal with chiral algebra ${\cal A}$)\foot{
One could also imagine imposing boundary conditions that
violate \Abcond\ for some of the generators of the chiral algebra.
Such boundary conditions can reduce, or at least modify the structure
of the chiral algebra on the half-space. A familiar, though simple,
version of this is the choice of Neveu-Schwarz or Ramond sectors in
the open string.}.

\newsec{Boundary integrable models}

The initial focus of this paper will be insertions of boundary
perturbations of the form:
\eqn\bdryact{ A_{bdry} ~=~ \mu~ \int_{-\infty}^{\infty} ~\chi(t) ~dt
\ , }
where $\chi(t)$ is some operator defined upon the boundary at $x=0$.
To be relevant (or marginal), the operator $\chi$ must have dimension
less than (or equal to) one.  For a relevant boundary operator, the
coupling constant, $\mu$, introduces a boundary mass scale.

I will also consider bulk perturbations of the form:
\eqn\bulkpert{ \Delta A_{bulk} ~=~ g ~ \int_{{\cal R}e(z) < 0}
 ~\psi(z,\bar z)  ~d^2 z \ , }
where $g$ is a coupling constant, and $\psi(z,\bar z)$ is generically
a sum of products of holomorphic and anti-holomorphic operators:
$\psi(z,\bar z) = \sum_j \psi_j(z) \tilde \psi_j(\bar z)$.

\subsec{Conformal perturbation theory}

To first order in conformal perturbation theory, the crucial issue in
whether a boundary perturbation leads to an integrable boundary field
theory is precisely what {\it representation} the boundary operator
has with respect to the underlying chiral algebra on the half-space,
and {\it not} how this operator may, or may not, be obtained from the
limit of bulk operators. This is because one tests a putative
conserved current, ${\cal O}(z)$, in the presence of perturbation,
$\psi(z,\bar z) = \psi(z) \tilde \psi(\bar z)$, by computing
corrections to the conformal Ward indentity $\bar \del {\cal O}(z) =
0$. This current will remain conserved if the corrections to this
Ward identity have the form $L_{-1} {\cal X}$, for some operator
${\cal X}$. At first order in perturbation theory, this will be true
if and only if the coefficient of the simple pole in the operator
product ${\cal O}(z) \psi(w)$ is $L_{-1}$ of some operator \Zamo. The
result of this operator product computation depends upon the operator
${\cal O}(z)$ and the representation generated by $\psi(w)$ under the
action of the chiral algebra (in particular, the null vectors are
usually crucial). If one finds that the coefficient of the simple
pole is indeed $L_{-1}$ of something for a particular bulk operator,
$\psi$, then this is a statement of representation theory, and must
also be true for a boundary operator in the same representation of
the chiral algebra of the theory on the half-space. However, for the
boundary theory, the operator $L_{-1}$ generates translations that
preserve the boundary, that is, $L_{-1} = {\del \over \del t}$.
Therefore, the Ward identity $\bar \del {\cal O}(z) = 0$ gets
corrected by ${\del \over \del t}$ of some boundary operator. Thus
one obtains the boundary corrections to a conserved charge in the
boundary integrable model in exactly the same manner that one
computes the bulk corrections to a conserved current in a bulk
integrable model. This means that, at least to first order in
perturbation theory, if one has a bulk perturbation that leads to a
quantum integrable field theory, then one can make a boundary
integrable field theory {\it provided} that one can construct a
well-defined boundary operator in the same representation of the
chiral algebra. The foregoing argument is basically a rephrasing of
the argument given in \GZam. The last proviso about actually
constructing boundary operators in the requisite representations is
where the subtleties lie \Cardy.

To illustrate these issues it is instructive to consider the Ising
model, for which there are two known bulk integrable models. The
simplest is obtained from the energy perturbation of the conformal
model, where the perturbing operator
belongs to the $\Phi_{1,3}$ representation of the Virasoro algebra.
This perturbation corresponds to giving a mass to the underlying
free fermion. Following the logic above, it follows that a
boundary operator in the $\Phi_{1,3}$ representation must lead to an
integrable model. This is indeed true \GZam, but the correponding
boundary operator is the boundary {\it magnetic } perturbation
\refs{\Cardy,\GZam}. The easiest way to see this is to consider the
Ising model with free boundary conditions, and look at what happens
when one takes a spin operator, $\sigma(z,\bar z)$, to the boundary.
At the boundary, the holomorphic and
anti-holomorphic parts of the spin field fuse into a single
representation of the Virasoro algebra. (Put differently, the
holomorphic part of the spin field meets its mirror image in the
boundary.) The net result
is to get the fusion product of two spin fields, which
generates the identity and the energy operator ($\Phi_{1,3}$). The
fusion to the identity disappears because one started with free
boundary conditions, and so one gets the required energy operator at
the boundary \Cardy. More generally, by making a choice of conformal
boundary conditions, one immediately restricts the representations of
operators that can be constructed on the boundary.

It is also instructive to consider the boundary analogue
of the other integrable perturbation of the Ising model. The bulk
perturbation corresponds to the magnetic operator, or the
$\Phi_{1,2}$ representation of the Virasoro algebra \Zamo. One cannot
obtain a boundary operator in this representation of the Virasoro
algebra by taking a limit of bulk operators. Indeed, if one starts
with the standard conformal boundary conditions (free or fixed) there
is simply no such operator in the boundary spectrum \Cardy. There
are, however, boundary condition changing operators in precisely the
right representation \Cardy, and so it is tantalizing to suggest that
a ``perturbation'' by such operators could yield the boundary
analogue of the bulk magnetic perturbation that leads to a quantum
integrable model with an $E_8$ structure \Zamo. The immediate problem
that one runs into is that one needs answer the question: what is the
conformal fixed point about which one is perturbing? Alternatively,
what is the U.V. fixed point of such a theory
where the boundary coupling vanishes? It is certainly not fixed or
free boundary conditions for the reasons mentioned above. One might
hope to find some more general conformal boundary
condition. Indeed, if one attempts a vague interpretation of what the
boundary condition changing operator in the $\Phi_{1,2}$
representation does, then it seems to want to make a U.V. fixed point
that would look like an average over a boundary magnetic field that
takes values $\pm \infty$ or $0$ for random intervals of random
length. It is, as yet, unclear to me whether this idea can be
turned into a more precise or meaningful proposal.

The essential conclusion is that perturbations of conformal field
theories fall into two classes, those that are in representations
of the chiral algebra that can be obtained in the boundary spectrum
of some conformal boundary condition, and those that are not.
If the bulk perturbation leads to a bulk integrable model, then
the former class of ``good'' perturbations lead to boundary
integrable models, while the latter class are ``problematic''
perturbations whose naive boundary analogues may not exist.

One should bear in mind that the
arguments above are only to first order in conformal perturbation
theory. If the perturbing operators are very relevant (\ie of low
dimension) then simple dimensional arguments can be used to show that
first order perturbation theory is sufficient to establish the
existence of higher spin conserved charges to all orders \Zamo. On
the other hand, if the perturbing operators are nearly marginal, then
there could be higher order corrections to the conserved charges. For
bulk perturbations, we know from free field formulations, and related
Toda theories, that even though there might be such corrections, they
do not spoil quantum integrability \Toda. Given the argument above,
and the close parallel between bulk and boundary operators, it seems
very plausible to expect the same to be true for boundary
perturbations\foot{There are however some complications if one
attempts
to use Toda and free field arguments directly for boundary
field theories. This is discussed below.}.

Suppose now that one makes simultaneous and
independent perturbations on the boundary and in the bulk using
\bdryact\ and \bulkpert, where $\chi(t)$ and $\psi(z)$ are in the
same representations of the chiral algebra. It follows from the first
order conformal perturbation theory that if a current ${\cal O}(z)$
leads to a conserved charge in the presence of the bulk perturbation,
then it also does so in the presence of the independent bulk and
boundary perturbations. It is, of course, quite possible that the
bulk and boundary perturbations could interact at higher orders in
perturbation theory and thus further constrain, or perhaps even
destroy, quantum integrability. Once again, this seems unlikely,
and if one considers perturbing operators of low enough dimension
then it cannot happen.

\subsec{Spin-$1$ currents and topological charges}

There are two important
exceptions to the standard first-order conformal perturbation theory
argument presented above. The first is trivial, and concerns
spin-$1$ currents. The holomorphic and anti-holomorphic parts, $J(z)$
and $\widetilde J(\bar z)$, of a spin-$1$ current in a conformal
model are separately conserved. If one makes a bulk perturbation,
then $J(z)$ and $\widetilde J(\bar z)$ give rise to a single
conserved $U(1)$ current if and only if the perturbing operator is
neutral with respect to some linear combination of the associated
left-moving and right-moving charges. While this is obvious, the
computation in terms of conformal perturbation theory is
substantially different from the argument above, and most
importantly, even though a bulk perturbation can preserve a $U(1)$
charge, the corresponding boundary perturbation may well destroy it
(even at first order).

The second exception is a little more subtle, and arises whenever
there are bulk topological charges. Such topological charges appear
in massive theories with extended supersymmetry. Even at first order
in perturbation theory one finds that a superconformal algebra will
receive corrections that generate first order corrections involving
total derivatives of bosonic fields. These corrections therefore
contribute bosonic terms to the algebra, and these depend upon the
boundary conditions. In infinite systems the boundary conditions are
usually fixed, and these boundary terms are referred to as
topological charges (for example, see \refs{\OW, \FMVW, \WLNW,
\ABL}).
In the semi-infinite system these topological charges will only be
conserved if the bosonic boundary terms are fixed. {\it A Priori}
this means that the
superalgebra will be, to some extent, broken unless some form of
Dirichlet boundary conditions are imposed. Therefore, even though
first order perturbation theory indicates no problems for each
supercharge individually, the supersymmetry can be broken even at
first order. I suspect that this can only occur in systems that have
conserved currents of fractional spin, whereas, in the more
conventional systems with commuting integer spin charges, the issue
of such topological charges will probably not arise.

For supersymmetric theories in semi-infinite systems, there several
ways in which one can ameliorate, or even avoid, the problems caused
by topological charges, and thereby preserve the supersymmetries.
One can:

\item{(i)}  Enforce Dirichlet boundary conditions.
\item{(ii)} Keep the bulk conformally invariant, and only perturb on
the boundary. There are no topological charges in a bulk
superconformal algebra.
\item{(iii)}  Find combinations of supercharges such that the
topological
charges do not effectively contribute to the commutator.
\item{(iv)}  Try to compensate for the topological charge terms
by using boundary degrees of freedom, and making re-definitions
of the boundary supersymmetry and hamiltonian.

To illustrate possibility (iv), suppose that a bulk topological
charge term appears in the square of a supersymmetry generator and
gives rise to a boundary topological term ${\cal X}$.  One can
cancel this term by introducing boundary fermions $b$ and
$b^\dagger$,
with $\{b,b^\dagger\} = 1$, and then adding boundary correction
of $b -  {\cal X} b^\dagger$ to the supercharge.  Indeed, it will
be shown in section 7 that it is precisely this mechanism that
enables one to construct $N=2$ supersymmetry boundary integrable
models with non-trivial boundary interactions.

\subsec{An Aside: Toda models}

In the presence of a boundary, the simply-laced classical Toda
models, of rank more than one, are more tightly constrained than the
perturbed conformal models:  In the Toda models, the boundary
couplings are fixed in terms of the bulk couplings
\refs{\durham,\itals}.  What is also surprising is that the
non-simply-laced models are not so  tightly constrained \durham.
Thus the Toda models all give rise to quantum integrable boundary
theories, but some have the boundary and bulk mass scales locked
together while others do not.

A correspondence between Toda models, with imaginary coupling
constants, and perturbed conformal field theories can be made by
viewing all but one, or two, of the Toda potential terms as defining
a Coulomb gas prescription, and then the remaining potential terms
are re-interpreted as free field forms of the relevant perturbations
of the conformal model \refs{\CGints,\Toda}. This correspondence has
proved a very valuable tool in bulk theories, and one would therefore
hope to be able to extend it to boundary theories. However, conformal
perturbation theory produces no constraints between bulk and boundary
mass scales.

There are thus several possible ways in which one might
account for this lack of constraint. First, it is possible that the
constraint appears at higher orders in conformal perturbation theory.
However, the exact $S$-matrices and effective actions suggest
otherwise; and in some instances one might be able to use dimensional
analysis to exclude the possibility of such higher order corrections.
Another possibility is that the introduction of massive, non-linear
boundary conditions at the level of the Toda action could conceivably
destroy, or interfere with the Coulomb gas interpretation,
particularly if the boundary terms modify the scaling properties of
what one hopes to interpret as the screening charges.  A third
possibility, which is supported by later results in this paper, is
that to properly define the boundary action so as to enable a Coulomb
gas interpretation, one may need to introduce new boundary degrees of
freedom and specify their dynamics. Without these degrees of freedom,
the Coulomb gas methods might fail, and the correspondence with
conformal perturbation theory may be be lost.

There is also an important observation that could provide some
explanation of the apparent difference between boundary Toda theories
and conformal perturbation theory in systems with boundary. Suppose
that a bulk perturbation of a conformal model were in a
representation of the chiral algebra that was not in the boundary
spectrum of any conformal boundary conditions for the bulk model --
just like the $\Phi_{1,2}$ perturbation of the conformal Ising model.
Then it will be impossible to find a boundary perturbation of the
corresponding Toda model (with imaginary coupling) such that the
model remains quantum integrable and possesses a limit in which the
bulk becomes massless, while the boundary remains massive. It follows
that there will be families of bulk perturbations that lead to bulk
integrable models, but that do not have boundary analogues for which
there is a massless bulk limit. It is natural to conjecture that the
models in which the bulk and boundary mass scales are fixed in terms
of one another are precisely in this class. Certainly the
perturbations of Ising fit this pattern.

If this conjecture is correct, it is then intriguing that one is, in
fact, able to construct a boundary Toda model with non-trivial
boundary interaction in the first place, since this would imply that
conformally forbidden boundary interactions will become permissible
so long as they are always combined with suitable massive bulk
interactions.

It is obviously valuable to check this conjecture against the
known examples.  It would also be useful
to see whether there is a characterization of this boundary spectrum
statement that can be made entirely in terms of a Toda field theory.
Since this issue is not central to this paper, I will not
pursue it further here.

\newsec{Some supersymmetry preserving boundary perturbations}

I will, for simplicity, focus on the $N=2$ superconformal minimal
models  with $A$-type modular invariants, and whose central charge is
$c = 3k/(k+2)$.  The relevant bulk perturbations of these models
that lead to $N=2$ supersymmetric integrable field theories are
well known \refs{\FMVW,\FLMW,\susyint}.  There are three distinct
such perturbations, but the one of importance here is the
Chebyshev perturbation.  That is, one perturbs the bulk using
\eqn\bulkp{ \lambda~\int~\Gminus \widetilde \Gminus ~\phi_k^+
(z, \bar z) ~d^2z ~+~ \bar \lambda~\int~\Gplus \widetilde \Gplus ~
\phi_k^- (z, \bar z) ~ d^2z \ ,}
where $\lambda$ is a coupling constant, $\phi_k^+$ is the
chiral primary field of charge $q = \tilde q = {k \over k+2}$
and conformal weight $h = \tilde h = {k \over 2(k+2)}$, and
$\phi_k^-$ is the anti-chiral conjugate of $\phi_k^+$.
This perturbation leads to a massive $N=2$ supersymmetric
integrable model whose effective \LG potential is the Chebyshev
polynomial of degree $(k+2)$ (see, for example \NWTrieste).
This may be thought of as the
$N=2$ superconformal analogue of the energy perturbation of the
ordinary minimal series.

Suppose that $\psi^\pm(t)$ are boundary operators that
transform in exactly the same representations as the
holomorphic operators $\Gmp ~\phi_k^\pm(z)$ under the action of
the $N=2$ superconformal algebra.  In particular this implies
that the operators $\psi^\pm(t)$ may be written
\eqn\psidefn{\psi^\pm(t) ~=~ \Gmp \Big(\hat\phi_k^\pm (t)\Big) \ ,}
where $\Gmp$ are the supercharges of the boundary theory, and
$\hat \phi^\pm(t)$ are boundary operators in the same
$N=2$ superconformal representation as $\phi^\pm (z)$ and
$\phi^\pm (\bar z)$.

In order to use $\psi^\pm(t)$ in a boundary perturbation it is
necessary to show that these operators are in the boundary spectrum
of some conformally invariant boundary condition. This is
straightforward, and is completely parallel with the analogous
situation in the Ising model. Let $\phi^+_1(z,\bar z)$ be the most
relevant chiral primary field (with $q = \tilde q = {1 \over k+2}$
and conformal weight $h = \tilde h = {1 \over 2(k+2)}$). This field
is the basic order parameter of the massive, bulk integrable model
\refs{\ntwoLG,\FMVW}. If we send this bulk field to the boundary, the
result is an operator in the fusion product of $\phi^+_1(z)$ with
itself. There are two fields in this fusion product: (i) $\phi^+_2$,
the chiral primary with $q = {2 \over k+2}$ and conformal weight $h =
{2 \over 2(k+2)}$, and (ii) the field $\Gplus \hat \phi^-_k$, with $q
= 1 - {k \over k+2} = {2 \over k+2}$ and conformal weight $h = \half
+ {k \over 2(k+2)} = {k+1 \over k+2}$. If one starts with the $N=2$
superconformal model with free boundary conditions in the \LG
formulation, then the expectation value of $\phi^+_2 = (\phi^+_1)^2$
will vanish at the boundary, and one will get the sub-leading
operator $\psi^-(t) = \Gplus \hat \phi_k^- (t) $. Similarly, one
obtains $\psi^+(t) = \Gminus \hat \phi_k^+ (t) $ by sending the
anti-chiral conjugate, $\phi_1^- (z, \bar z)$, of $\phi_1^+ (z, \bar
z)$, to the boundary.

Conformal perturbation theory can now be invoked to show that
(at least to first order) one can obtain a boundary integrable model
from a simultaneous bulk perturbation using \bulkp\ and boundary
perturbation using \bdryact\ with:
\eqn\chidefn{\chi(t) ~=~ \nu ~\psi^+(t) ~+~ \bar \nu ~\psi^-(t) \ .}
The constant $\nu$ is a complex phase, with complex conjugate $\bar
\nu$. If there is no bulk perturbation then one can absorb this phase
$\nu$ into a re-definition of $\phi_k^\pm$. If there is a bulk
perturbation, then this freedom of re-definition can be used to
adjust
the phase of $\lambda$ (in \bulkp) or the phase $\nu$. Thus there are
three independent couplings: the bulk mass scale, determined by
$|\lambda|$, the boundary mass scale, determined by $\mu$ in
\bdryact, and the relative phase between $\lambda/\bar \lambda$ and
$\nu/\bar \nu$.

It is well known that the bulk perturbation \bulkp, in the infinite
domain, preserves all of the supersymmetry \refs{\FMVW,\PFKI,\LNW},
and provides one with a massive $N=2$ supersymmetric model. There are
two ways of seeing the latter: One can either explicitly verify the
result using conformal perturbation theory, or, more generally, one
can appeal to the theory of supersymmetric actions, which states that
perturbations that involve only top components of superfields (as
does \bulkp) preserve the supersymmetry. The same arguments can also
be naively applied to the boundary perturbation \chidefn, with
similar conclusions. However, the massive bulk model has topological
charges, and so the supersymmetry variation of the bulk action
generates boundary terms via total derivatives. As described earlier,
unless one chooses Dirichlet boundary conditions, or make the bulk
massless, one is going to have to be very careful with the boundary
interaction if one is going to preserve $N=2$ supersymmetry. I will
return to this issue later, but first I think it useful to
present a simple and fairly complete example, and re-examine the
whole issue of supersymmetry using exact scattering matrices.

\newsec{The sine-Gordon model}

The simplest example of an $N=2$ superconformal minimal model is
the $k=1, c=1$ minimal model that can be realized by a single free
boson compactified at the ``supersymmetric'' radius.  The
superconformal generators can be written in terms of the holomorphic
part of a canonically normalized boson, $\varphi(z)$, as:
\eqn\susybose{  T(z) ~=~ -\coeff{1}{2}~ (\partial \varphi(z))^2 \ ;
 \qquad J(z) ~=~  \coeff{i}{\sqrt{3}}~ \partial \varphi(z) \ ;
\qquad G^\pm(z)  ~=~  e^{\pm i \sqrt{3} \varphi(z)} \ .}
If $\tilde \varphi(\bar z)$ denotes the anti-holomorphic part of the
boson, then the $N=2$ superconformal boundary condition implies that
$\varphi(z)|_{x=0}  = \tilde \varphi(-\bar z)|_{x=0}$.  The order
parameter, and its conjugate,  are given by $\phi_1^\pm(z, \bar z) =
e^{\pm {i \over \sqrt{3}} (\varphi(z) + \varphi(\bar z))}$,
and as $x \to 0$ this
becomes $e^{\pm{2i \over \sqrt{3}} \varphi(z)} |_{x=0} = \Gmp
\phi_1^\pm(z)|_{x=0}$.  The bulk integrable perturbations are
$\Gmp \widetilde \Gmp \phi^\pm_1 =  e^{\mp {2i \over \sqrt{3}}(
\varphi(z) + \varphi(\bar z))}$.   We are thus dealing with the
boundary sine-Gordon theory described of \GZam, with action:
\eqn\bdrySG{ \eqalign{\int_{-\infty}^{\infty} \int_{x<0}~ \big(
\coeff{1}{2} (\partial_t \Phi)^2 - & \coeff{1}{2} (\partial_x
\Phi)^2 \big) ~-~ {M \over \beta}~ \big[ cos \big(  \beta \Phi
\big) ~-~ 1 \big] ~dx~dt \cr
 ~-~ & {2m \over \beta} ~ \int_{-\infty}^{\infty}~ cos
\coeff{\beta}{2} \big(  \Phi -  \Phi_0 \big) ~dt \ .}}
{}From this one finds that $\Phi$ satisfies the usual sine-Gordon
equation with boundary condition:
\eqn\sGbcs{\partial \Phi \big |_{x=0}  ~=~ m~sin \coeff{\beta}{2}
\big(  \Phi -  \Phi_0 \big) \big |_{x=0} \ .}
The boson, $\Phi$, has the standard normalization of
sine-Gordon theory, for which the supersymmetry point
corresponds to $\beta^2 = {16 \pi \over 3}$.  The parameters
$M, m$ and $\Phi_0$ coincide with the three parameters
described in the last section.  It is known that this model
is indeed integrable for arbitrary choices of $M, m$ and $\Phi_0$,
and so the boundary and bulk perturbations are simultaneously
and independently integrable.

\subsec{Supersymmetry}

The bulk $S$-matrix for the sine-Gordon model has been known for a
long time \ZandZ. At the supersymmetry point one is in the repulsive
regime, and so the fundamental spectrum consists of the two kinks:
a soliton and an anti-soliton; there are no breathers.
The soliton and anti-soliton have fermion numbers
$+\half$ and $-\half$ respectively, and they form a two-dimensional
supermultiplet \refs{\FMVW, \ABL, \PFKI}.  As explained in \ABL, the
bulk $S$-matrix commutes with the generators of the affine quantum
group in the principal gradation.  At the supersymmetry point, the
affine quantum group generators have spin $\half$ and are precisely
the four generators of the massive superalgebra. On the doublet
consisting of the soliton and anti-soliton, the action of these
supersymmetry generators is given by:
\eqn\SAsusy{ {\cal Q}_\pm(\theta) ~=~ C ~e^{+ {\theta \over 2}} ~
E_\pm ~ q^{\pm { H \over 2} } \ ; \qquad \qquad  \widetilde
{\cal Q}_\pm (\theta) ~=~
C ~e^{- {\theta \over 2} } ~ E_\pm ~ q^{\mp {H \over 2}} \ , }
where $H$, $E_+$ and $E_-$ are the $2 \times 2$ matrices generating
the $SU(2)$ algebra, with $[H,E_\pm] = \pm 2 E_\pm$ and
$[E_+,E_-] = H$. (I will use the notation and conventions
of \ABL.)  In \SAsusy, the parameter, $\theta$, is the rapidity of
the kink that is being acted upon, and $C$ is a constant that
satisfies $C^2 = {|\lambda|\over 2 \pi i}(1 - q^{2}) $.
The parameter, $q$, is the usual parameter of the quantum group,
and is given by $q = -e^{ i \pi (1 - 8 \pi/\beta^2)}$, which
reduces to $q = i$ at the supersymmetry point.

To extend the action of the quantum group onto multi-kink states
one introduces the co-product $\Delta$.
At the supersymmetry point, the co-product introduces phases
appropriate to ``commuting'' fermionic generators through kinks of
fermion number $\pm \half$.  The fact that the supersymmetries
commute with the $S$ matrix is then written:
\eqn\Scomm{\big[{\cal S}(\theta) \ , \ \Delta({\cal Q}_\pm(\theta))
\big] ~=~ \big[ {\cal S}(\theta) \ , \ \Delta( \widetilde
{\cal Q}_\pm (\theta))\big] ~=~ 0 \ . }

The algebra of the bulk supercharges is:
\eqn\supalg{\eqalign{\big\{{\cal Q}_\pm  \ , \  \widetilde
{\cal Q}_\pm \big\} ~=~ & 0 \ \ ; \qquad \big\{{\cal Q}_\pm \ ,\
\widetilde {\cal  Q}_\mp \big\} ~=~  \coeff{|\lambda|}{2 \pi i}~
\big( 1~ - ~ q^{\pm 2 {\cal T}} \big) \cr  \big\{{\cal Q}_+ \ ,\
{\cal  Q}_- \big\} ~=~  & \coeff{|\lambda|}{2 \pi i}~q^{-1}~
(1 - q^2)~ {\cal P}_+ \ \ ; \quad\ \big\{ \widetilde
{\cal Q}_+ \ ,\  \widetilde{\cal  Q}_- \big\} ~=~
\coeff{|\lambda|}{2 \pi i}~  q~(1 - q^2)~ {\cal P}_-  \ ,} }
where $q = i$.   The operators $\coeff{|\lambda|}{2 \pi i}
{\cal P}_\pm$ are the light-cone components of the momentum,
and ${\cal T} = H$, is the generator of the $U(1)$ in
$U_q(SU(2))$, and is given by the bulk topological charge:
\eqn\topchg{{\cal T} ~=~ H ~=~ \coeff{\beta}{2 \pi}~
\int_{-\infty}^{\infty} ~\partial_x \Phi ~dx \ .}
For a bulk superconformal model, the
coupling, $\lambda$, and momenta, ${\cal P}_\pm$, are scaled in such
a manner that $\coeff{|\lambda|}{2 \pi i} {\cal P}_\pm$ remains
finite, while the topological terms: $\coeff{|\lambda|}{2 \pi i}~
(1 - q^{\pm 2 {\cal T}})$ scale to zero.  As a result, there are
no topological terms in the algebra in the bulk massless limit.

\subsec{Incorporating the boundary}

For the boundary theory \bdrySG, conformal perturbation
theory leads one to expect that two supersymmetry generators will
survive.  There are two possibilities for the pair of
surviving supercharges:
\eqn\bdryQ{{\rm either} \qquad \widehat {\cal Q}_\pm ~=~
{\cal Q}_\pm ~+~ e^{\pm i \omega}~q^{-1} ~\widetilde {\cal Q}_\pm
\qquad  {\rm or} \qquad \widehat{\cal Q}^\prime_\pm ~=~
{\cal Q}_\pm ~+~ e^{\pm i \omega}~q^{-1}~\widetilde{\cal Q}_\mp \ .}
The phase, $\omega$, is an arbitrary constant.
The choice between $\widehat {\cal Q}_\pm$ and $\widehat
{\cal Q}_\pm^\prime$ depends upon the boundary condition for
the $U(1)$ current.  The algebras are slightly different:
\eqn\Qalg{\eqalign{ \big\{ \widehat {\cal Q}_\pm  \ , \
\widehat {\cal Q}_\pm \big\} ~=~ & 0 \ \ , \cr
\big\{ \widehat  {\cal Q}_\pm \ ,\ \widehat {\cal  Q}_\mp
\big\} ~=~ & \coeff{|\lambda|}{2 \pi i}~q^{-1}~
(1 - q^2)~\Big[({\cal P}_+ + {\cal P}_- ) ~+~
\cr & \qquad \qquad  \coeff{1}{(1 - q^2)}~  \big(
2 cos \omega - e^{ -i \omega} q^{+ 2 {\cal T}} - e^{i \omega}
q^{- 2 {\cal T}}  \big) \Big] \ ;}}
whereas, one has:
\eqn\Qpalg{\eqalign{ \big\{ \widehat {\cal Q}^\prime_\pm  \ , \
\widehat {\cal Q}^\prime_\pm \big\} ~=~ & 2 ~e^{\pm i \omega}
{}~q^{-1}~\coeff{|\lambda|}{2 \pi i}~ \big( 1~ - ~ q^{\pm 2
{\cal T}} \big)\ \ , \cr  \big\{ \widehat  {\cal Q}^\prime_\pm
\ ,\  \widehat  {\cal  Q}^\prime_\mp  \big\} ~=~ & q^{-1}~
\coeff{|\lambda|} {2 \pi i}~(1 - q^2)~({\cal P}_+ ~+~
{\cal P}_- )  \ .}}
Note that only the combination, ${\cal P}_+ + {\cal P}_-$,
appears in these algebras. Indeed, the relative phases in \bdryQ\
were chosen so as to produce this combination.
This is because the boundary conserves energy, but not momentum,
and the operator ${\cal P}_+ + {\cal P}_-$ is
the dimensionless energy operator: $E/M_0$, where $M_0$ is the
soliton mass.

While both the superalgebras, \Qalg\ and \Qpalg, have two generators,
they are different. In the first, one cannot separate the topological
terms from the energy operator, whereas one can in the second
algebra. One can also see this distinction is not an artefact of
basis choice by looking at unitarity bounds on the energy that
can be deduced from the superalgebra.

To compute precisely what is happening with supersymmetry in this
model one can use the exact boundary reflection matrix of \GZam. That
is, {\it if one assumes that there is no boundary structure}, one can
use boundary Yang-Baxter, crossing, unitarity and bootstrap, to
arrive at the amplitudes for reflection of kinks off the boundary.
For the soliton/anti-soliton doublet, this amplitude can can be
written \GZam:
\eqn\bdryref{{\cal R}(\theta) ~=~ X(\theta, \xi, k) ~ \left(
\matrix{cos(\xi - i \lambda \theta)  & - \coeff{k}{2}
{}~sin(2 i \lambda \theta) \cr - \coeff{k}{2}~sin(2 i
\lambda \theta) & cos(\xi + i \lambda \theta)} \right) \ ,}
where $k$ and $\xi$ are parameters, and $\lambda \equiv
{8 \pi \over \beta^2} - 1 = {1 \over 2}$ at the supersymmmetry
point \foot{Do not confuse this parameter, $\lambda$, with that
of \bulkp.}.  The parameter $\xi$ is simply related to $\Phi_0$
in \bdrySG\  \refs{\GZam, \SSW}:
\eqn\xiphi{\xi~=~ {4\pi \over \beta}~\Big(\Phi_0 ~-~ {2\pi \over
\beta}~ {\rm Int } \Big[{\beta\Phi_0 \over 2\pi} ~+~ \half \Big]
\Big) \ .}
The parameter $k$ is a complicated function of $\Phi_0$ and the
boundary mass $m$. The Dirichlet boundary conditions, $\Phi|_{x=0} =
\Phi_0$, corresponds to setting $k=0$, while Neumann boundary
conditions correspond to taking $\xi = \Phi_0 = 0$, and $k = [sin(\pi
\lambda/2)]^{-1} = \sqrt{2}$ at the supersymmetry point \GZam. In the
massless bulk limit, \FSW, the parameter $\xi$ (or $\Phi_0$) becomes
a trivial (and irrelevant) shift of the free boson $\Phi$, and the
reflection matrix becomes:
\eqn\Rmassless{{\cal R}(\theta) ~=~ \widetilde X(\theta - \theta_B)~
\left(\matrix{e^{\lambda (\theta - \theta_B)/2}  & i e^{- \lambda
(\theta - \theta_B)/2 } \cr i e^{- \lambda (\theta - \theta_B)/2}
& e^{\lambda (\theta - \theta_B)/2}} \right) \ ,}
where $\theta_B$ is a parameter related to the boundary mass scale.
For Dirichlet boundary conditions, $\theta_B \to -\infty$, while
for Neumann boundary conditions, $\theta_B \to +\infty$.

To have supersymmetry in the exact quantum model, the supercharges
must ``commute'' with the reflection matrix.  More precisely,
one should be able to act with a supersymmetry before or after
reflection, and obtain the same result.  This means that we must
seek a linear combination, ${\cal Q}(\theta)$, of the supersymmetry
generators that satisfies:
\eqn\QandR{ {\cal Q}( \theta) ~ {\cal R} (\theta) ~=~
{\cal R} (\theta) ~ {\cal Q}(- \theta) \ .}
The minus sign on the right-hand-side comes from the fact that a
kink of rapidity $\theta$ reflects as a kink of rapidity $-\theta$.
For general values of $k$ and $\xi$, there is exactly one
solution to this equation:
\eqn\onesusy{{\cal Q}(\theta)  ~=~ e^{+ i \xi}~ \big(
{\cal Q}_+ ~+~ ~ q^{-1}  ~\widetilde {\cal Q}_- \big) ~+~
e^{ - i \xi}~ \big(  {\cal Q}_- ~+~  ~ q^{-1} ~\widetilde
{\cal Q}_+ \big) \ . }
For Dirichlet boundary conditions one finds two supercharges:
\eqn\twosusy{\widehat {\cal Q}_\pm (\theta)  ~=~
{\cal Q}_\pm ~+~ q^{-1}~ e^{\mp 2 i \xi}~\widetilde {\cal Q}_\pm \ ,
}
and these have eigenvalues $\pm 2$ with respect to the
(conserved) topological charge, ${\cal T}$.
When the bulk is massless one still generically finds only one
supersymmetry, and this is given by \onesusy\ with $\xi = 0$. If one
also imposes Dirichlet boundary conditions one finds two
supersymmetries, given by \twosusy\ with $\xi = 0$. For Neumann
boundary conditions (with a massless bulk) one also finds two
supersymmetries, and these are:
\eqn\neumsusy{\widehat {\cal Q}^\prime_\pm(\theta)  ~=~
{\cal Q}_\pm ~+~ q^{-1} ~\widetilde {\cal Q}_\mp \ . }
It should also be noted that these two supersymmetries still
commute  with the reflection matrix \bdryref\ in the $k \to
\infty$ limit even when the bulk is massive.

\subsec{Multi-solitons states and the co-product}

While I have identified certain linear combinations of supercharges
that commute appropriately with the boundary reflection matrix, there
is an intrinsic danger that such linear
combinations of quantum group generators may not be respected
by the co-product, and so the symmetry algebra may not be
well-defined on multi-particle states.

The action of the co-product on the supersymmetry generators \SAsusy\
is:
\eqn\coprod{\eqalign{\Delta\big( {\cal Q}_\pm \big) ~&=~
{\cal Q}_\pm(\theta_1) \otimes 1 ~+~ q^{\pm H} \otimes {\cal Q}_\pm
(\theta_2) \cr  \Delta\big(\widetilde {\cal Q}_\pm \big) ~&=~
\widetilde {\cal Q}_\pm(\theta_1) \otimes 1 ~+~ q^{\mp H} \otimes
\widetilde {\cal Q}_\pm (\theta_2) \ ,}}
where $\theta_j$ is the rapidity of the $j^{\rm th}$ kink.  One sees
that the linear combination \neumsusy\ transforms consistently
under the co-product, whereas the combinations \twosusy\ and
\onesusy\
do not:  they get different factors of $q^{\pm H}$ for
different terms in the linear combination.  Thus, the terms that make
up the conserved charge will be weighted by phases that
depend upon the total topological charge of the state.
It is perfectly reasonable for the form of a conserved fractional
spin
charge to depend upon the total topological charge of the
state, and this will not cause a problem provided the topological
charge
be conserved.  Certainly the bulk $S$-matrix conserves topological
charge,
but the reflection matrix \bdryref\ violates topological charge,
except at the Dirichlet point.  This violation is always
in multiples of $\pm 2$, and the {\it relative} phases in the
co-product are $q^{\pm 2 H}$  (\ie from $q^{\pm  H}$ to $q^{\mp H}$),
and so the violation of topological charge by boundary reflection
only involves factors of $q^{\pm 4}$.  At the supersymmetry point
this is not a problem, since $q^{\pm 4} = 1$.

The conclusion is that all the supercharges described in the previous
section extend consistently to multi-particle states in the presence
of
the boundary.  However, if one wishes to consider the more general
Coulomb gas, with arbitrary values of $q$, then the analogue of
\onesusy\ will not be a symmetry of the theory.  Indeed, one will
only
have a consistent fractional spin symmetry either at
at the Dirichlet point, where topological charge is conserved, or
where the boundary reflection matrix is anti-diagonal, and there
are two symmetries given by \neumsusy.

\subsec{Bogolmolnyi bounds}

Since the exact quantum theory described by the structureless
reflection matrix of \GZam\ still has $N=1$ supersymmetry, one can
use this to place bounds upon the energy of the ground state.

{}From the positivity of the expectation value of
$\{{\cal Q}^\dagger ,  {\cal Q}\}$, one finds that
the following operator must have non-negative expectation
value:
\eqn\poscomm{\coeff{|\lambda|}{2 \pi } ~\left[ ({\cal P}_+
+ {\cal P}_- ) ~+~ \left( 2 cos (2 \xi) - e^{ +2 i \xi} q^{+ 2
{\cal T}} - e^{-2 i \xi} q^{- 2 {\cal T}}  \right) \right] \ .}
If the topological charge, ${\cal T}$, is not conserved
then the ${\cal T}$ dependent terms in the foregoing must
be interpreted as part of the Hamiltonian, and one simply
finds that the hamiltonian has non-negative expectation value.
If, however, one has Dirichlet boundary conditions, the
ground state will have topological charge given by \topchg.
If $\Phi \to 0$ as $x \to -\infty$, then the topological charge
of the ground state is ${\cal T}_0 = \coeff{\beta}{2 \pi} \Phi_0$.
{}From \xiphi\ one sees that $ \xi  = \coeff{8 \pi^2}{\beta^2}
{\cal T}_0$ (provided $- \pi < \beta \Phi_0 < \pi$), and using
 $q = -e^{ i \pi (1 - 8 \pi/\beta^2)} = e^{ - i 8 \pi^2/\beta^2}$,
one obtains $q^{\pm 2 {\cal T}_0} =  e^{\mp 2i\xi}$.
Using this in \poscomm\ one obtains a lower bound on the energy
of the ground state:
\eqn\Ebound{ E/M_0 ~\ge~ sin^2(\xi) ~=~
sin^2(\coeff{8 \pi^2} {\beta^2} {\cal T}_0) ~=~
sin^2(\coeff{3 \pi}{2} {\cal T}_0 )\ ,}
where $M_0$ is the mass of the soliton.

It is amusing to compare this with the classical result.  If one
seeks the lowest energy configuration with $\Phi \to 0$ as
$x \to - \infty$, and $\Phi|_{x=0} = \Phi_0$, one finds that it
is a section of the single soliton solution, and if one computes
the energy, one finds $E = M_0 sin^2( {\pi \over 2} {\cal T}_0)$,
where ${\cal T}_0 = {\beta \Phi_0 \over 2 \pi}$  and
$M_0 = {8 M\over \beta^2}$ is the classical soliton mass.

The bound \Ebound\ is saturated if and only if the supercharge
annihilates the vacuum state. For a real supersymmetry it is quite
possible to get dynamical supersymmetry breaking \Wita, and thus this
bound may not be saturated. In real supersymmetric \LG theories, for
example, one finds that one can always formally solve
${\cal Q} | 0\! > = 0$, but the solution is not a normalizable
wave-function when the superpotential is odd, or equivalently, when
the index of the theory is even. On the other hand, if the model
possesses a complex, or holomorphic, $N=2$ supersymmetry, then the
solutions to ${\cal Q}_\pm | 0\! >= 0$ are always normalizable. The
bulk sine-Gordon model has a complex $N=2$ supersymmetry, and
for Dirichlet boundary conditions, this complex structure
is preserved\foot{This will be discussed further in section 7.},
and the supersymmetry will not be broken.  It is somewhat less
clear whether the $N=1$ supersymmetry of the more
general boundary conditions will be broken.

\subsec{Comments}

For the boundary reflection matrix \bdryref, I have shown that
two supersymmetries are preserved either when one imposes
Dirichlet boundary conditions, or when the bulk is massless and one
imposes Neumann boundary conditions.  Otherwise, the supersymmetry is
reduced to $N=1$.
These results are in apparent conflict with those of section 4, which
indicate that one should be able to preserve $N=2$ supersymmetry
for non-trivial boundary conditions.  While this
conflict might be partially resolved by the effects of boundary
contributions of topological charges, there is still a conflict
in the limit where the bulk is massless.  In this limit there
are no topological charges,  and yet \bdryref\ still gives rise to
only $N=1$ supersymmetry.

The key to understanding this discrepancy is a primary
assumption in the derivation of the boundary reflection matrix \GZam,
and in the imposition of \QandR.  That is, to arrive at these results
it was assumed that the boundary had no structure.  In terms of
scattering theory, this means that the boundary has no stable states,
and transforms trivially under all the symmetry algebras.  In terms
of field theory, the concept is rather more vague, but essentially it
is supposed to conjure the idea that either the boundary action
contains no new degrees of freedom, or, if the boundary action does
have new degrees of freedom, their evolution will be trivial in that
it is completely determined by the evolution of the boundary values
of bulk fields.  At any rate, the amount of supersymmetry will
obviously depend upon the boundary action, and, as we will see,
certain very particular choices lead to $N=2$ supersymmetry, while
others lead generically to $N=1$ supersymmetry.

\newsec{$N=2$ supersymmetric boundary reflection matrices}

The simplest method of introducing non-trivial boundary
structure given a known structureless exact reflection matrix,
$R^b_a(\theta)$, is to formally glue a particle to the
boundary\foot{I am grateful to P.~Fendley for explaining this
process to me.} \refs{\GZam,\Ghosh}.
To be explicit, one thinks of the new boundary as a combination
of the known boundary and a particle of formal rapidity $\zeta$
running parallel to the boundary (see \lfig\figtwo).  The rapidity
of the boundary particle is formal since this particle is never
considered as hitting the boundary -- the particle is simply
``glued'' to the boundary forever.  Let $\alpha, \beta, \dots$
denote the species of boundary particle, and let $a, b, c, \dots$
denote the species of bulk particle.  The new boundary reflection
matrix is then given by
\eqn\Rtilde{\widetilde R^{b \beta}_{a \alpha} (\theta) ~=~
S^{c \gamma}_{a \alpha} (\theta - \zeta) ~R^d_c (\theta) ~
S^{\beta b}_{\alpha d} (\theta + \zeta) \ .}
The matrix $\widetilde R^{b \beta}_{a \alpha}$ can now be viewed
as the reflection matrix for a boundary that has states $\alpha,
\beta, \dots$ (see \lfig\figtwo).  This matrix satisfies boundary
Yang-Baxter, unitarity, crossing and bootstrap as a consequence of
the fact that $S^{b \beta}_{a \alpha}$ and $R^b_a$ satisfy such
conditions.  Note that $\widetilde R^{b \beta}_{a \alpha}$ has one
more parameter, $\zeta$, compared to the original structureless
reflection matrix. The idea now is to consider the boundary
sine-Gordon model at the special points where there is $N=2$
supersymmetry. At such points there are no free parameters in
$R^b_a$, but the decorated reflection matrix of \Rtilde\ will have
exactly one parameter, which can be interpreted as the boundary mass
scale. Moreover, since each term on the right-hand-side of \Rtilde\
commutes with the two supersymmetry generators, the resulting
boundary reflection matrix will do so as well. In the process of
checking the commutation of $\widetilde R^{b \beta}_{a \alpha}$ with
the supersymmetry, one of course finds that the boundary transforms
as a supersymmetry doublet of rapidity $\zeta$.

This process can obviously be generalized by gluing more particles
to the boundary, and in principle there will be a parameter for
each particle.  In practice, once one fuses these boundary particles
into higher multiplets, one might find some redundancy.  For
simplicity,
I will only consider the gluing of a single particle here.

\figinsert\figtwo{Obtaining boundary reflection matrices by gluing
a particle of rapidity $\zeta$ to the boundary.}
{2.0in}{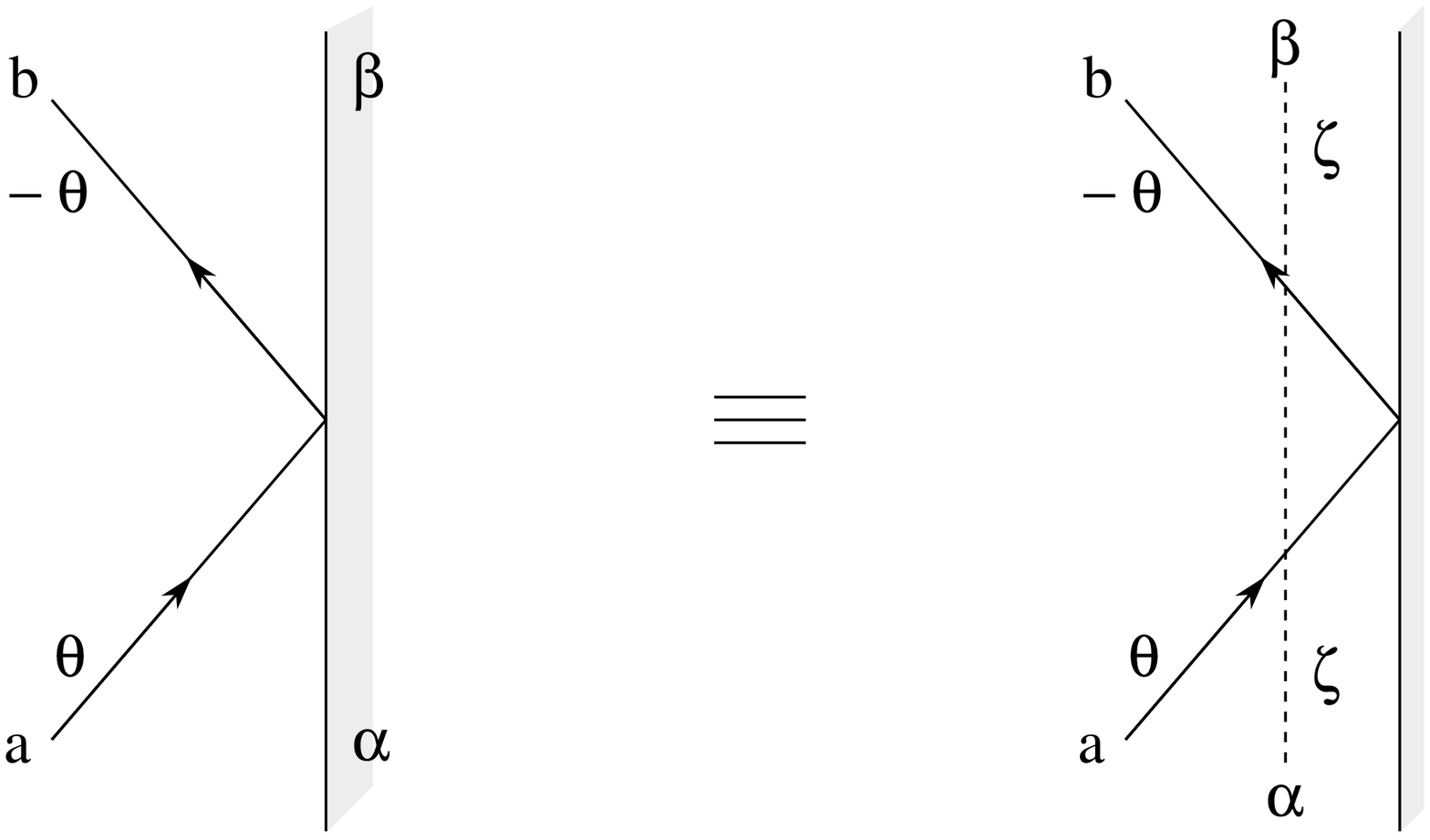}

\subsec{Reflection matrices preserving topological charge}

Consider first the structureless reflection matrix \bdryref\ with
$k=0$.  The bulk $S$-matrix has the form:
\eqn\bulkS{\eqalign{ a(\theta) ~\equiv~ S^{+ +}_{+ +}(\theta )~&=~
S^{- -}_{- -} (\theta ) ~=~ Z(\theta) ~ sin (\lambda(\pi + i \theta))
 \ , \cr b(\theta)  ~\equiv~ S^{+ -}_{+ -} (\theta )~&=~
S^{- +}_{- +} ~=~ - Z(\theta) ~ sin (i \lambda \theta) \ , \cr
c(\theta) ~\equiv~ S^{+ -}_{- +} (\theta ) ~&=~  S^{- +}_{+ -} ~=~
Z(\theta) ~ sin (\lambda \pi ) \ ,}}
where $\lambda = {8\pi \over \beta^2} -1$, and $Z(\theta)$ is
a well known normalization factor \ZandZ.
{}From this one obtains a reflection matrix with the following
non-vanishing terms:
\eqn\DirRtilde{\eqalign{\widetilde R^{\pm \pm}_{\pm \pm}
(\theta )~&=~ Y(\theta, \xi,\zeta) ~a(\theta + \zeta) ~ a(\theta
- \zeta) ~cos(\xi \mp i \lambda \theta ) \ , \cr \widetilde
R^{\pm \mp}_{\pm \mp}  (\theta )~&=~ Y(\theta, \xi,\zeta) \big[ ~
b(\theta + \zeta)~b(\theta - \zeta) ~ cos(\xi \mp i \lambda
\theta ) ~+~ \cr  & \qquad \qquad \qquad \qquad \qquad \qquad
c(\theta + \zeta)~ c(\theta - \zeta) ~ cos(\xi \pm i \lambda \theta )
\big] \ , \cr \widetilde R^{\pm  \mp}_{\mp \pm} (\theta ) ~&=~
Y(\theta, \xi,\zeta)  \big[ ~b(\theta + \zeta)~c(\theta -  \zeta) ~
cos(\xi \pm i \lambda  \theta ) ~+~ \cr  & \qquad \qquad \qquad
\qquad \qquad \qquad  b(\theta - \zeta)~ c(\theta + \zeta) ~
cos(\xi \mp i \lambda \theta ) \big] \ ,}}
where $Y(\theta, \xi,\zeta) = Z(\theta + \zeta) Z(\theta - \zeta)
X(\theta, \xi, k = 0)$.
One can now explicitly verify that, provided that one has $q^4 =1$,
this reflection matrix commutes appropriately with the action of two
supercharges of the form \twosusy\ but with the sign of the
second term in \twosusy\ reversed.  The reason for this sign reversal
and the constraint $q^4 = 1$ is precisely because one needs
consistency with the co-product as described in section 5.3.

\subsec{Reflection matrices from anti-diagonal scattering}

The other way to get $N=2$ supersymmetry is to use the generators
\neumsusy, which commute with \bdryref\ in the $k \to \infty$
limit.  The bulk does not need to be massless for this to be true,
but in the massless limit this corresponds to Neumann boundary
conditions.  I will first consider the more general problem with a
massive bulk.

The result of the fusion procedure outlined above is a reflection
matrix whose non-zero entries are:
\eqn\NeumRtilde{\eqalign{\widetilde R^{\mp \pm}_{\pm \pm}
(\theta )~&=~ \hat Y(\theta, \zeta) ~a(\theta - \zeta) ~ b(\theta
+ \zeta) \ , \cr \widetilde  R^{\pm \mp}_{\pm \pm}  (\theta )~&=~
\hat Y(\theta, \zeta) ~a(\theta - \zeta) ~  c(\theta +  \zeta) \ ,
\cr \widetilde  R^{\pm \pm}_{\pm \mp}  (\theta )~&=~
\hat Y(\theta, \zeta) ~a(\theta + \zeta) ~ c(\theta - \zeta)\ , \cr
\widetilde R^{\pm  \pm}_{\mp \pm} (\theta ) ~&=~ \hat Y(\theta,
\zeta)  ~a(\theta + \zeta) ~  b(\theta -  \zeta)  \ ,}}
where $\hat Y(\theta, \zeta)  = lim_{k \to \infty}
[- {k \over 2}  sin(2 i \lambda \theta) X(\theta, k, \xi)
Z(\theta + \zeta)  Z(\theta - \zeta)]$.

It should be noted that are no conserved $U(1)$ charges for either
this reflection matrix, or for the associated bulk scattering
problem.  Given the $k \to \infty$ limit of \bdryref\ one might
have hoped that there would be a conserved ``axial'' $U(1)$:
the soliton number of the right-movers {\it minus} the soliton
number of the left-movers.  However, this charge is violated
by the off-diagonal elements, $c(\theta)$ in \bulkS, in the
scattering of a left-mover off a right-mover.  This violation
is then transmitted through to \NeumRtilde\ in the fusion
procedure.  A $\ZZ_4$ subgroup of the $U(1)$ does survive, since
the axial soliton number is conserved mod $4$.  Since kinks have
fermion number $\pm \half$, this means that the reflection matrix
\NeumRtilde\ only violates fermion number mod $2$:  it therefore has
the physically desirable property of reflecting bosonic states as
bosonic states and fermionic states as fermionic states (unlike the
general form of \bdryref).  This symmetry also implies that the
non-vanishing reflection matrix elements \NeumRtilde\ all have an
odd number of $+$ indices  (or $-$ indices).

On the other hand, it is precisely the off-diagonal terms
in left-right scattering that vanish when one takes the
massless limit of the bulk scattering problem. In this limit,
one sends $\theta, \zeta \to \pm \infty$ (where the sign is
the same as the sign of $\theta, \zeta$ respectively).  One takes
this limit in such a manner that $\theta - \zeta$ remains finite, and
as a result the terms  $R^{\pm \mp}_{\pm \pm}$ of the boundary
reflection matrix \NeumRtilde\ vanish.  There is thus a conserved
charge: the sum of the bulk axial $U(1)$ charge and the boundary
charge.

In this massless limit one must also be a little careful in taking
the infinite rapidity limit of the supersymmetry generators \SAsusy.
If $\theta > 0 $ then ${\cal Q}_\pm$ will survive, and $\widetilde
{\cal Q}_\pm$ will go to zero; mutatis mutandis for $\theta < 0$.
This is simply the statement that in the conformal limit, $\widetilde
{\cal Q}_\pm$ must act trivially on right movers, while ${\cal
Q}_\pm$ will act trivially on the left movers. Another consequence of
this is that the anti-commutator $\{{\cal Q}_\pm, \widetilde {\cal
Q}_\mp \}$ in \supalg\ will vanish, as it must because of the absence
of bulk topological charges. As a result, the first anti-commutator
in \Qpalg\ will also vanish in this limit.

Finally, one should note that, for the reasons outlined in
section 5.3,  unlike the boundary reflection matrix
\DirRtilde, the matrix \NeumRtilde\ will commute with the
quantum group generators \neumsusy\ for arbitrary values of $q$.
Therefore, this formulation of the boundary interaction will
generalize to arbitrary Coulomb gas problems.

\newsec{Manifestly supersymmetric actions for systems
with boundary}

The issue of what kind of boundary conditions preserve how much
supersymmetry in a manifestly bulk supersymmetric theory has been a
recurrent question over the years. An associated issue is the kind of
boundary conditions needed to preserve a bulk Nicolai mapping (see,
for example, \HLuck). My intent here is to go further than simple
boundary conditions, and look at quantum mechanical systems on the
boundary that can be coupled to the bulk theory in such a manner
that supersymmetry is preserved. These models will be motivated by
the \LG formulations of the integrable theories described in
section 4.

The first step to constructing a manifestly supersymmetric
action is to make the elementary observation that
in such an action, the boundary perturbations, defined by
\psidefn\ and \chidefn, must be fermionic operators, and so
there is no way that they can be added by themselves to this
action.  The only option is to introduce a dimensionless boundary
fermion, $b$, and use fermion bilinears.
There are now two choices, $b$ can be real, or complex,
with either $\{b,b\} = 1$ {\it or}   $\{b,b\} =  \{b^\dagger,
b^\dagger\} = 0$, and $\{b,b^\dagger\} = 1$.   Motivated by the fact
the bulk massless limit of the models of interest have a
conserved $U(1)$ charge,  I will take $b$ to be complex. One
therefore needs to consider boundary perturbations of the form:
\eqn\goodbpert{\nu~ b^\dagger~\psi^+ ~+~ \bar\nu~\psi^-~b \ . }
Note that if $\psi^\pm$ transforms under a $U(1)$ charge, then
one can arrange that the action preserve that $U(1)$ by
making $b$ and $b^\dagger$ transform appropriately.

A more standard way to think of the boundary operators, $b$ and
$b^\dagger$, is as introducing a boundary spin degree of freedom
exactly as one does in the Kondo problem. Indeed, the quantization
rules for $b$ and $b^\dagger$ imply that they generate some (possibly
time dependent) representation of the gamma matrix algebra. In the
sine-Gordon model, the boundary interaction \goodbpert\ corresponds
to coupling vertex operators to spin raising and lowering operators.

For the sine-Gordon model, one can also see rather explicitly
that one should add these complex boundary degrees of freedom. One
considers the model as one approaches the limit of free boundary
conditions. As discussed in \GZam, any boundary  degrees of freedom
will not disappear in the free limit, but they
will simply decouple from the bulk and will become massless
boundary  excitations.  These will appear as poles at
$\theta = i \pi/2$ in the boundary reflection matrix.
One indeed finds such poles in all channels of the reflection
matrix, that is, in all the amplitudes: $S \to S$, $A \to A$,
$S \to A$ and $A \to S$, where $S$ and $A$ denote the soliton and
anti-soliton respectively \foot{The poles in the off-diagonal part
of the $S$-matrix were not so much of a concern in \GZam\ since
the focus was mainly on values of the coupling for which these
poles are cancelled by zeroes in the numerator.  At the
supersymmetry point there are no such cancellations.}.
Since the soliton and the anti-soliton have fermion
numbers $\pm \half$, the off-diagonal poles indicate that there
will be boundary excitations that carry fermion number $\pm 1$.
The corresponding operators are $b$ and $b^\dagger$.

To go much further requires one to consider more specific,
manifestly supersymmetric models.   The problem is then a common
one in conformal field theory:  there are no manifestly
conformal actions for the models considered in section 4.
Instead, I will start with the simplest superconformal model
for which we have such an action, and then move on to
$N=2$ supersymmetric \LG models, which provide
good descriptions of the models of section 4.  This will
enable me to obtain the exact boundary actions and
boundary superpotentials corresponding to the boundary
reflection matrices of section 6.2.

\subsec{Non-trivial, supersymmetric boundary interactions}

Consider the free $N=2$ superconformal model with a complex
boson and a complex Dirac fermion.  This model has central charge
$c=3$.  The (Euclidean) action will be taken to be:
\eqn\freeact{\eqalign{\int_{- \infty}^0 ~dx
\int_{-\infty}^{\infty}~dy & \big[ - (\partial_x \phi)
(\partial_x \bar \phi) ~-~ (\partial_y \phi)(\partial_y \bar
\phi) ~+~  \coeff{i}{2}(\bar \lambda \gamma^\mu \partial_\mu
\lambda - (\partial_\mu \bar \lambda) \gamma^\mu  \lambda))
\big] \cr &~+~ \int_{-\infty}^{\infty}~  i \Big(b^\dagger
{d \over dy} b \Big) ~-~ \shalf  (\bar \lambda  \gamma^*
\lambda)\big|_{x=0}~dy ,}}
where $x = x^0$, $y = x^1$, $\bar \lambda$ is the hermitian
conjugate of $\lambda$, and
$$\gamma^0 = \left(\matrix{0 &i \cr-i &0 \cr} \right)  \qquad
\qquad \gamma^1 = \left(\matrix{1 &0 \cr 0 &-1 \cr} \right)
\qquad \gamma* = -i \gamma^0 \gamma^1 =
\left(\matrix{0 &1 \cr 1 &0 \cr} \right) \ .$$
A boundary term has already been added to the standard free
action in \freeact.  This is motivated by the boundary action
for the Ising model (see, for example, \GZam), and it incorporates
the free boundary fermions.  The boundary term involving
$\lambda$ has been included to simplify
computations, and in particular, it leads naturally
to the free boundary conditions for the bulk fermion.  One should
also note that without this $\bar \lambda \gamma^* \lambda$ term,
the action is  invariant under both the left and right handed
fermion number symmetries independently, whereas one knows that
only one linear combination of these $U(1)$ symmetries will survive
on the half-space.   The boundary term has effectively selected which
$U(1)$ symmetry survives, and this is the axial $U(1)$:
\eqn\axuone{\lambda_j ~\to~ e^{i \nu} \lambda_j \ ,  \qquad
\bar \lambda_j ~\to~ e^{- i \nu} \bar \lambda_j \ ,}
where $\lambda_j$ are the components of $\lambda$.

The variation of \freeact\ yields the bulk
equations of motion, along with the boundary terms:
\eqn\varonb{\eqalign{- \int_{-\infty}^{\infty}~ \big[
(\partial_x \bar \phi) & (\delta \phi) ~+~ (\partial_x  \phi)
(\delta \bar \phi) \big]\big|_{x=0} ~-~ i \Big[ \Big({d \over dy}
b^\dagger \Big) ~ \delta b ~+~ \Big({d \over dy} b \Big)~
\delta b^\dagger \Big] \cr ~+~ & \shalf\big[(
\bar \lambda_1 - \bar \lambda_2) (\delta \lambda_1 + \delta
\lambda_2) ~-~ (\lambda_1 - \lambda_2) (\delta \bar \lambda_1 +
\delta\bar \lambda_2)\big]\big|_{x=0} ~dy \ .}}
Requiring that \varonb\ vanish imposes Neumann,
Dirichlet, or a combination of Neumann and Dirichlet
conditions on the real and imaginary parts of the boson,
along with analogous boundary conditions on the fermion.

The bulk part of the action \freeact\ also has an $N=2$
superconformal symmetry, which can be written in terms of
components as:
\eqn\susytrf{\eqalign{\delta \phi ~&=~ -(\lambda_1 \bar \alpha_1 +
\lambda_2 \bar \alpha_2)\ , \qquad \qquad \delta \bar \phi ~=~ (\bar
\lambda_1  \alpha_1 + \bar \lambda_2 \alpha_2) \ , \cr
\delta \lambda_1 ~&=~ -[\partial_x \phi - i \partial_y \phi]~
\alpha_2 \ , \qquad \qquad \delta \bar \lambda_1 ~=~ -[\partial_x
\bar \phi + i \partial_y \bar \phi]~ \bar \alpha_2 \ , \cr
\delta \lambda_2 ~&=~ [\partial_x \phi + i \partial_y \phi]~
\alpha_1 \ ,  \qquad \qquad \delta \bar \lambda_2 ~=~ [\partial_x
\bar \phi  - i \partial_y \bar \phi]~ \bar \alpha_1 \ ; }}
where $\alpha_1, \alpha_2, \bar \alpha_1$ and $\bar \alpha_2$
are the complex fermionic supersymmetry parameters.  If one uses
this in the action \freeact, and assumes that $\delta b =
\delta b^\dagger = 0$, then one generates the following
boundary terms:
\eqn\susyvar{\eqalign{ - \shalf  \int_{-\infty}^{\infty}~ \big[
& (\partial_x \phi) (\bar \lambda_1 - \bar \lambda_2)  ~-~ i
(\partial_y \phi)  (\bar \lambda_1 +  \bar \lambda_2) \big]
(\alpha_1 - \alpha_2) \cr
& ~-~  \big[(\partial_x \bar \phi) (\lambda_1 - \lambda_2)  ~-~
i (\partial_y  \bar \phi)  (\lambda_1 +  \lambda_2) \big]
(\bar \alpha_1 - \bar \alpha_2) \ .}}

To preserve the supersymmetry in the system with boundary, one
must choose boundary conditions that cause \varonb\ and \susyvar\
to vanish\foot{It is in making these choices that one ordinarily
selects which $U(1)$ fermion number symmetry will be preserved in
the boundary theory.}. Further restrictions might follow from
requiring that
the boundary conditions are consistent with \susytrf.
Obvious solutions are:

\item{(i)}  Dirichlet boundary conditions for
the bosons, $\phi|_{x=0}  = const$, along with $\alpha_1 =
\alpha_2 = \alpha$.  Requiring consistency with \susytrf\ means
that we must also impose $(\lambda_1 + \lambda_2)|_{x=0} =0$.
\item{(ii)} Neumann boundary conditions for
the bosons, $\partial_x \phi|_{x=0} = 0$, along with $\alpha_1
= \alpha_2 = \alpha$.  Requiring consistency with \susytrf\ means
that we must also impose $(\lambda_1 - \lambda_2)|_{x=0} =0$.
\item{(iii)} A combination of Dirichlet and Neumann conditions, in
which one takes $\partial_x (\phi + \bar \phi)|_{x=0} = 0$, and
$(\phi - \bar \phi)|_{x=0} = const$.  One then finds that one
must impose $(\lambda_j + \bar \lambda_j)|_{x=0} = 0$, and
$\alpha_j = \bar \alpha_j$, $j =1,2$.

The first two choices preserve a complex $N=2$ superconformal
algebra, with parameters $\alpha$ and $\bar \alpha$, while the
last alternative is apparently a real $N=2$ algebra, but one
can find a new basis in which it will be a complex supersymmetry.

The boundary perturbation analogous to \goodbpert\ is given by
the same expression, but with
$\psi^\pm$ replaced by $(\lambda_1 + \lambda_2) \big({\partial^2
V \over \partial \phi^2} \big)$ and $(\bar \lambda_1 + \bar
\lambda_2) \big({\partial^2 \bar V \over \partial \bar \phi^2}
\big)$ respectively, where $V$ is some superpotential
function.  Supersymmetry then requires a boundary action
of the form:
\eqn\dbdryact{\shalf~\int_{-\infty}^\infty ~ \Big({\partial^2
V \over \partial \phi^2} \Big) ~ b^\dagger ~(\lambda_1 +
\lambda_2) ~-~ \Big({\partial^2 \bar V \over \partial \bar
\phi^2} \Big)~b~(\bar \lambda_1 + \bar \lambda_2)  ~+~ \Big|
{\partial V \over \partial \phi} \Big|^2 ~dy\ .}
Combining the variation of this with \varonb, one finds that
the equations of motion of the boundary degrees of freedom
are:
\eqn\bdryeqn{\eqalign{(\lambda_1 - \lambda_2)|_{x=0}  ~&=~ b~\Big(
{\partial^2 \bar V \over \partial \bar \phi^2} \Big) \Big|_{x=0}\ ;
\qquad    {d \over dy}~b ~=~ -\coeff{i}{2}~(\lambda_1 +
\lambda_2)~\Big({\partial^2  V \over \partial \phi^2} \Big)
\Big|_{x=0} \ ; \cr
(\partial_x \phi)|_{x=0} ~&=~\shalf \Big({\partial V \over \partial
\phi}  \Big) \Big({\partial^2 \bar V \over \partial \bar \phi^2}
\Big) \Big|_{x=0} ~- ~b~(\bar\lambda_1 + \bar \lambda_2)~ \Big(
{\partial^3 \bar V \over \partial \bar \phi^3} \Big)\Big|_{x=0}
\ ;}}
along with the complex conjugates of these equations.

This model is also invariant under a complex $N=2$
supersymmetry.  All one needs to do is set $\alpha_1 =
\alpha_2 = \alpha$ in \susytrf, and use \susytrf\ along with:
\eqn\deltab{ \delta b ~=~  -\Big({\partial V \over \partial \phi}
\Big) \Big|_{x=0}~ \alpha \ ; \qquad \delta b^\dagger ~=~
- \Big({\partial \bar V \over \partial \bar \phi} \Big)\Big|_{x=0}~
\bar \alpha \ .}
This $N=2$ supersymmetry is consistent with $U(1)$ symmetry \axuone\
provided that the boundary fermions and supersymmetry parameters
transform according to:
\eqn\axact{\alpha ~\to~ e^{i \nu} \alpha \ ,  \qquad
\bar \alpha ~\to~ e^{-i \nu} \bar \alpha \ , \qquad
b ~\to~ e^{ i \nu} b \ , \qquad
b^\dagger ~\to~ e^{ -i \nu} b^\dagger \ .}

\subsec{Introducing a bulk superpotential}

Consider now what happens when the bulk becomes a massive \LG theory.
That is, temporarily ignore all boundary terms, and add the following
to the bulk Euclidean action:
\eqn\bulkpot{\int_{- \infty}^0 ~dx \int_{-\infty}^{\infty}~dy~
\bigg[ \Big({\partial^2 W \over \partial \phi^2} \Big)~\lambda_1
\lambda_2 ~-~\Big({\partial^2 \overline W \over \partial\bar
\phi^2}  \Big)  ~\bar \lambda_1 \bar \lambda_2 ~-~
\Big|{\partial W \over \partial \phi}  \Big|^2 \bigg] \ ,}
for some scalar superpotential, $W$.  To preserve the bulk
supersymmetry one must add the following terms to the
supersymmetry transformations of the fermions:
\eqn\Wsusy{\delta_W \lambda_i ~=~ \epsilon^{ij}~\Big({\partial
\overline W \over \partial\bar \phi} \Big)~ \bar \alpha_j \ , \qquad
\qquad  \delta_W \bar \lambda_i ~=~ \epsilon^{ij}~\Big({\partial
W \over \partial \phi} \Big)~ \alpha_j \ ; }
where $\epsilon^{ij} = -\epsilon^{ij}$ and $\epsilon^{12} = +1$.
This bulk action preserves the vector $U(1)$ symmetry:
\eqn\vecuone{\lambda_1 ~\to~ e^{i \nu} \lambda_1 \ ,  \qquad
\lambda_2 ~\to~ e^{-i \nu} \lambda_2 \ ; \qquad
\bar \lambda_1 ~\to~ e^{ i \nu} \bar \lambda_1 \ , \qquad
\bar \lambda_2 ~\to~ e^{- i \nu} \bar \lambda_2 \ ,}
with the supersymmetry parameters transforming in a similar manner
(but with the subscripts $1,2$ interchanged).

If $W$ is quasi-homogeneous, that is, if
\eqn\quasihom{W(a^\omega \phi) ~=~ a W(\phi) \ ,}
for some $\omega$ and any value of $a$, then this
action also preserves an $R$-symmetry.  Namely,
it is invariant under:
\eqn\Rsymm{\eqalign{\phi ~&\to~ e^{i \omega \nu}~\phi \ , \qquad
\bar \phi ~\to~ e^{-i \omega \nu}~\bar \phi \ , \cr \lambda_j ~&\to~
e^{-{i \over 2} (1 - 2 \omega)  \nu}~\lambda_j \ , \qquad \bar
\lambda_j ~\to~ e^{+{i \over2} (1 - 2 \omega) \nu}~\bar \lambda_j
\ , \qquad j = 1,2 \cr \alpha ~&\to~  e^{-{i \over 2} \nu}~\alpha \ ,
\qquad \bar \alpha ~\to~ e^{+{i \over2}\nu}~\bar \alpha \ .}}

Now consider the boundary theory.  There are two different ways of
getting $N=2$ supersymmetry, and these correspond
to the two distinct $N=2$ algebras described earlier.  One can
take:
\eqn\susych{{\rm (i)} \quad \alpha_j = \bar \alpha_j\ , j =1,2
\qquad {\rm or} \qquad {\rm (ii)} \quad \alpha =  \alpha_1 =
\alpha_2\ ,\quad  \bar \alpha = \bar \alpha_1 =
\bar \alpha_2 \ .}
The second choice is the more interesting topologically since the
supersymmetry transformations involving $W$ and those involving
derivatives of fields remain independent.  Amongst other things, this
is crucial  to relating the $N=2$ supersymmetric theory to
a topological \LG theory \topLG.  The first choice is more akin
to  a real supersymmetry.  One can also verify directly from the
supercurrents \refs{\WLNW} that the first superalgebra has the
form \Qalg, whereas the second choice leads to a bulk superalgebra
of the form \Qpalg.  Indeed, the topological charge for the
\LG model is \refs{\OW,\FMVW}:
\eqn\Wtopch{{\cal T} ~=~ \int ~ \partial_x W ~dx \ ,}
and, for example, one can write the supercurrent that generates
the transformation with parameter $\alpha$ as:
\eqn\supcur{{\cal G}_+ ~=~ \bar \lambda_1 ~(\partial_x \phi +
i \partial_y \phi) ~-~ \bar \lambda_2 ~(\partial_x \phi -
i \partial_y \phi) ~+~ (\lambda_2 - \lambda_1) \Big({\partial W
\over \partial \bar \phi} \Big) \ . }
Here I am taking the convention that the short-distance expansion of
the fermions is normalized to $\bar \lambda_1(z) \lambda_2 (w) \sim
{1 \over z-w}$ and $\bar \lambda_2(\bar z) \lambda_1 (\bar w) \sim
{1 \over \bar z - \bar w}$. Taking the operator product of
${\cal G}_+$ with itself generates simple pole terms whose net
contribution to the anti-commutator of this supercharge with itself
is  $\int_{-\infty}^0~ 4 \big({\partial W \over \partial \phi} \big)
(\partial_x \phi)  dx =  4 W|_{x=0}$.

To make the boundary theory consistent, one must once again
ensure that all the boundary variations vanish, and then check
consistency with the combination of \susytrf\ and \Wsusy.
The ordinary variation of \bulkpot\ does not give any new
boundary terms in \varonb, but the supersymmetry variation of
\bulkpot\ does give rise to the following additional boundary term:
\eqn\Wvaronb{  \shalf  \int_{-\infty}^{\infty}~ \bigg[
\Big({\partial W \over \partial \phi} \Big)~(\lambda_1 +
\lambda_2) (\alpha_1 + \alpha_2)  ~-~  \Big({\partial \overline W
\over \partial \bar \phi} \Big)~(\bar \lambda_1 + \bar \lambda_2)
(\bar \alpha_1 + \bar \alpha_2) \bigg]\bigg|_{x=0} ~dy\ .}
Suppose for the moment that boundary potential $V$ to zero.
If one looks for supersymmetry of type (ii), then
to remove the terms in \Wvaronb\ it suffices to set either
$(\lambda_1 + \lambda_2)|_{x=0} = 0$ or ${\partial W \over \partial
\phi}\big|_{x=0} = 0$.  However, requiring that
\varonb\ vanish and checking consistency with \susytrf\ and \Wsusy\
leads to the Dirichlet conditions:
$\phi_{x=0} = \phi_0$ and $(\lambda_1 + \lambda_2)|_{x=0} = 0$.
Note that these boundary conditions break the vector-like
$U(1)$ symmetry, but if $W$ is quasi-homogeneous, then they
preserve the $R$-symmetry.

If one looks for a {\it purely} Neumann boundary conditions, then
the  supersymmetry transformations lead one to impose:
\eqn\Neum{\eqalign{\alpha_1 ~&=~ \alpha_2  ~=~ \bar \alpha_1 ~=~
\bar \alpha_2 \ ; \qquad (\lambda_1 -  \lambda_2)|_{x=0}~=~0 \ ; \cr
\partial_x \phi |_{x=0} ~&=~ \Big({\partial
\overline W \over  \partial \bar \phi} \Big)
\Big|_{x=0}  \ ;  \qquad \partial_x \bar \phi |_{x=0} ~=~ \Big(
{\partial W \over  \partial \phi} \Big)\Big|_{x=0} \ .}}
Note that the first of these equations means that the supersymmetry
has been reduced to $N=1$.  The constraints \Neum\ do
not guarantee the vanishing of \Wvaronb.  To
accomplish this one needs to add another obvious boundary term:
\eqn\bdryW{\int_{-\infty}^{\infty}~(W(\phi) ~+~ \overline W
(\bar \phi))|_{x=0}  ~dy \ .}
This extra boundary term is also crucial for the bosonic
boundary conditions in \Neum\ to follow from
the variation of the action.
This term is also nothing other than the boundary
contribution to the bulk topological charge \Wtopch.  It is also
amusing to note that for Dirichlet boundary conditions, where the
topological charge is conserved, one has the option of adding
\bdryW\ since it will make no difference to the boundary variation.
Thus, when the topological charge is conserved, one can chose
whether, or not, to include such terms in the action, and hence
hamiltonian.  When the topological charge is not conserved, one no
longer has the option, and one {\it must} include such the
topological charge terms in the action.   At the same
time, one also sees a reduction from complex $N=2$ to real $N=1$
supersymmetry.  These boundary conditions also break all $U(1)$
symmetries, reducing any $R$-symmetry to a discrete subgroup.

If one now seeks $N=2$ supersymmetry of type (i), one
is led to a combination of Neumann and Dirichlet
conditions.  One uses an action that includes \bdryW, and one
takes:
\eqn\mixed{\eqalign{ \alpha_j ~&=~ \bar \alpha_j \ ; \qquad
(\lambda_j
+ \bar \lambda_j)|_{x=0} ~=~ 0 \ ;  \qquad j = 1,2 \ ; \cr
(\phi - \bar \phi)|_{x=0} ~&=~ 2i \phi_0 \ ;
\qquad [\partial_x(\phi + \bar \phi)]|_{x=0} ~=~ \Big(
{\partial W \over  \partial \phi}  ~+~ {\partial \overline W \over
 \partial \bar \phi}\Big)\Big|_{x=0} \ .}}
Note that the non-trivial boundary dynamics only involve the
real part of $\phi$, and so the boundary ``quantum mechanics''
is that of a single real scalar field.  These boundary conditions
break any $R$-symmetries, but preserve the vector-like $U(1)$
symmetry \vecuone.  (The boundary term in \freeact\ that breaks
this $U(1)$ vanishes identically as a consequence of the boundary
conditions.)

One can now combine the foregoing with a non-trivial boundary
interaction, that is, one considers an action that is the sum of
\freeact, \dbdryact, \bulkpot\ and \bdryW.  This action only
possesses an $N=1$ supersymmetry.  However, exactly as in section 6,
one can adjust the non-trivial boundary dynamics so as to get a
type (ii) $N=2$ supersymmetry.  One takes an action that is the
sum of \freeact, \dbdryact\ and \bulkpot, but does {\it not}
include \bdryW.  One then cancels the variation \Wvaronb\ by the
simple expedient of setting:
\eqn\VandW{\Big( {\partial V \over  \partial \phi} \Big)
\Big|_{x=0} ~=~\mu~ W|_{x=0} \ ,}
and changing \deltab\ to:
\eqn\newdeltab{ \delta b ~=~ 2 \bar \mu^{-1}\bar \alpha ~-~ \mu~
\alpha ~W|_{x=0}\ ; \qquad \delta b^\dagger ~=~ 2 \mu^{-1}
\alpha ~-~ \bar \mu~\bar \alpha~\overline W|_{x=0}~  \ .}
The parameter $\mu$ is an arbitrary complex number whose modulus
can be thought of as the ratio of the boundary and bulk mass scales.
The real and imaginary parts of $\mu$, along with the bulk mass
scale are precisely the three independent parameters identified
in the conformal perturbation theory of section 4.
For the sake of completeness, the classical equations of motion
of this $N=2$ supersymmetric model are:
\eqn\ELeqns{\eqalign{(\lambda_1 - \lambda_2)|_{x=0} ~&=~ 2
\bar \mu~b~ \Big( {\partial \overline W \over  \partial \bar \phi}
\Big) \Big|_{x=0} \ , \ \quad {d \over dy}~b ~=~ \coeff{i}{2}~\mu~
(\lambda_1 + \lambda_2)~\Big({\partial W \over \partial \phi} \Big)
\Big|_{x=0} \ , \cr (\bar \lambda_1 - \bar \lambda_2) |_{x=0} ~&=~2
\mu~b^\dagger~\Big( {\partial W \over  \partial  \phi} \Big)
\Big|_{x=0} \ , \ \quad {d \over dy}~b^\dagger ~=~ \coeff{i}{2}~
\bar \mu~(\bar \lambda_1 + \bar \lambda_2)~\Big({\partial \overline W
\over \partial \bar \phi} \Big) \Big|_{x=0} \ , \cr  \partial_x
\phi |_{x=0} ~&=~ - \bar \mu~b~(\bar \lambda_1 + \bar \lambda_2)
\Big({\partial^2 \overline W \over \partial \bar \phi^2} \Big)
\Big|_{x=0} ~+~ 2~|\mu|^2 ~W~\Big({\partial \overline W \over
\partial \bar \phi} \Big) \Big|_{x=0} \ .}}
One can easily check that these are consistent with the
type (ii) $N=2$ supersymmetry transformations.

Having found a combined boundary and bulk interaction that
preserves $N=2$ supersymmetry it is useful to observe that one
can now add more species of boundary fermion, with whatever
potentials one desires and still keep $N=2$ supersymmetry.
Specifically, one can add any number of boundary fermions
with couplings of the form \dbdryact, and with supersymmetry
transformations  given by \deltab.  The choice of bosonic
potentials for these extra fermion species is also arbitrary.
Presumably only a small subset of these models will be quantum
integrable.

\subsec{Comments}

There are several important things to note about this  last result.
First, in order to get $N=2$ supersymmetry, the boundary and bulk
superpotential must be related by \VandW.  This should not be a
surprise given the discussion in sections 3 and 4, where the boundary
integrable perturbation is obtained directly from the bulk integrable
perturbation. The derivative in \VandW\ may be a little surprising,
but is a natural consequence of the fact that the boundary
perturbation, defined by \psidefn\ and \goodbpert, only involves the
action of one supercharge whereas the bulk perturbation, \bulkp,
involves the action of two supercharges. It is also amusing to note
that for the bulk Chebyshev perturbation, the superpotential, $W$, is
a polynomial of degree $k+2$. This means that the bulk bosonic
potential has $k+1$ zero-energy minima, while the boundary bosonic
potential has $k+2$ zero-energy minima.  The constraint between
boundary and bulk potential is also familiar in sine-Gordon and
Toda theories.  One should note, however,
that here the bulk and boundary mass scales are independent

The next thing to observe is that the boundary topological charge
terms \bdryW\ have been done away with completely.  What has
happened is that the new transformation \newdeltab\ has
introduced the following boundary terms into the supercurrents:
$$ {\cal G}_{b,+} ~=~  2 \mu^{-1} b ~-~ \mu~b^\dagger~ W|_{x=0} \ ;
\qquad {\cal G}_{b,-} ~=~ 2 \bar \mu^{-1} b^\dagger
{}~-~ b~\bar \mu~ \overline W|_{x=0} \ .$$
If one computes the anti-commutator each of these charges with
itself, then one generates either $- 4 W|_{x=0}$ or $- 4 \overline
W|_{x=0}$, which are precisely what one needs in order to cancel the
boundary contribution of the topological terms coming from the
anti-commutators of the bulk supercharges. (See the discussion
following \Wtopch.)  Thus one can reverse the foregoing perspective
entirely, and take the view that the problem of boundary
contributions
of bulk topological charge terms has been solved by introducing
$W$-dependent terms into the supersymmetry variation of the
boundary fermions, and this, in turn requires the boundary
dynamics described by \dbdryact\ and \VandW.

One should also observe that because one has managed
to do away with the terms \bdryW, the action will now preserve
the $R$-symmetry when superpotential is quasi-homogeneous.
Otherwise this $U(1)$ is broken generically to a $\ZZ_2$,
(the fermion number mod $2$), or perhaps a larger discrete
group if the superpotential has additional symmetry.

The $N=2$ supersymmetric boundary \LG theories constructed here
clearly provide the effective field theories of the perturbed
models described in section 4.  This correspondence is well
established for the bulk, but here I have been able to get the
exact effective boundary potential from the knowledge of the bulk
theory and the existence of $N=2$ supersymmetry.
The fact that one can preserve the $R$-symmetry of the
model when the superpotential is quasi-homogeneous not only
leads to the scale invariance of the interaction, and to
an all important conserved $U(1)$ current, but it also means
that one should be able to formulate a Coulomb gas description
of this boundary model based upon the methods of \refs{\FGLS,\Witb,
\EGenNW}.

The structure of this effective theory is also completely parallel
to the scattering theory described in section 6.2.  If the model is
massless in the bulk then there is always $N=2$ supersymmetry, along
with a conserved axial $U(1)$ symmetry.  If the bulk is massive,
then the axial symmetry is broken down to conservation of the
fermion number mod 2, and the supersymmetry is reduced to $N=1$.
However, if one chooses exactly  the right {\it non-trivial} boundary
dynamics then one can restore  the $N=2$ supersymmetry (but not the
axial $U(1)$).  Moreover, the process of getting ever more
complicated
boundary reflection matrices via the gluing procedure must coincide
with
the introduction of more and more boundary
fermion species in the \LG formulation. Indeed, introducing
additional
species of boundary fermion also extends the boundary Hilbert
space by a spin doublet for each such fermion.  One can then extract
higher spin boundary states by projecting onto irreducible factors in
the tensor product of the doublets.  This is very reminiscent of the
the Kondo problem, which can described by a  resonant level
model \PWAF\ in which the spin impurity is created using boundary
fermions. The Kondo problem can also be described in terms of exact
$S$-matrices on a system with a boundary, with the spin impurity
being described in terms a boundary potential when the impurity
is underscreened, or in terms of kinks when the impurity is
overscreened \PFKondo.

Finally, there is something of a conundrum with the type (i)
$N=2$ supersymmetry of \mixed.  The indications from
sections 5 and 6 are that there should be an exactly solvable
model with $N=2$ supersymmetry, and a conserved vector-like
charge.  It seems very plausible that the effective
action for this model is the one with a structureless boundary
that leads to \mixed.  However, it does not appear to be
possible to implement this form of $N=2$ supersymmetry in the
presence of non-trivial boundary interactions.  Thus
the effective actions corresponding to the boundary reflection
matrices described in section 6.1 are, as yet, unknown.

\newsec{Conclusions}

In this paper it has been shown, using both field theory and
exact $S$-matrix theory,  that there are quantum integrable,
boundary field theories that have non-trivial boundary dynamics
and possess both $N=1$ and $N=2$ supersymmetry.
This was done initially by using conformal perturbation theory, and
these results were further substantiated by the construction of the
manifestly supersymmetric \LG effective actions that had all the
properties that one would expect based upon conformal perturbation
theory and the known results of bulk \LG theory.
It was also shown that one could easily construct $N=1$
supersymmetric actions with independent boundary and
bulk actions, but one only gets $N=2$ supersymmetry when the
boundary action is related to the bulk action in exactly the manner
suggested by the conformal perturbation theory.

A directly parallel story occurs with the construction of
the exact boundary reflection matrices.  The structureless
reflection matrices for the sine-Gordon model only have $N=1$
supersymmetry, except at special points in the parameter space,
where the supersymmetry is
enhanced to $N=2$.  This coincides with the results coming from
simplest boundary interactions in the effective \LG theory.
In order to get $N=2$ supersymmetric exact reflection matrices,
one can go to these special $N=2$ supersymmetric points for the known
structureless reflection matrix, and
then construct more complicated such matrices by decorating
the boundary by gluing kinks to it.  One then finds natural
candidate reflection matrices for the integrable models
described by the field theory.  An issue that is not fully
resolved here is precisely how many glued kinks correspond to
the basic $N=2$ supersymmetric \LG model with only one species of
boundary fermion.  One way to resolve this is to apply the ideas of
\PFKondo, and perform a Bethe Ansatz computation to determine the
boundary entropy, and then use this to determine the number of wells
in the boundary potential.  This is presently under investigation.

{}From the results of section 7.1, one sees that one can couple a
free
$N=2$ supersymmetric theory to an arbitrary boundary superpotential,
and still preserve $N=2$ supersymmetry.  However, based upon
the results of section 7.2, I do not believe that this will
remain true in any $N=2$ superconformal field theory.  To get a
specific non-trivial $N=2$ superconformal model from an  effective
action, one needs to use a specific quasi-homogeneous superpotential.
This means that the boundary potential is fixed by the $N=2$
supersymmetry.  Similarly for perturbed $N=2$ supersymmetric boundary
field theories:  the effective boundary potential for the quantum
integrable perturbation is uniquely determined by the choice of the
superconformal model, and the corresponding bulk integrable
perturbation, even when the bulk remains massless.   These potentials
are thus  known explicitly by virtue of the supersymmetry and the
known form of the integrable bulk superpotentials.  If one wants to
have more exotic boundary interactions one must introduce more
fermionic boundary degrees of freedom.

The results presented in this paper are also highly encouraging
for the use Coulomb gas methods for more general boundary integrable
models.  In sections 5 and 6, I identified two natural superalgebras
on the half-space, and observed that one of them only had a well
defined action at the supersymmetry point, whereas the other would
obviously generalize to other values of the quantum group
parameter, $q$.  It also turned out that the latter ``good''
superalgebra was the one that coincided with the superalgebras that
came from conformal perturbation theory and effective \LG actions.
As a result, I expect that the more general conformal perturbation
theory and Coulomb gas ideas should be generalizable to boundary
theories.  I also anticipate that the results will be qualitatively
similar to those of section 7, in that one will have to use
appropriate {\it non-trivial} boundary degrees of freedom, and that
the boundary theory will probably have some qualitative multi-well
effective potential that depends upon the value of $q$, the quantum
group parameter\foot{H.~Saleur has also come to similar conclusions
by other methods, and I am grateful to him for conversations on this
issue.}.

The fact that one can very probably generalize Coulomb gas methods
to the half-space, suggests that, if one introduces the proper
boundary degrees of freedom, one should be able to get the
correspondence between conformal field theories and Toda models
to work in systems with boundary as well.  Thus any discrepancy
between the two approaches will occur either because one has not
used exactly the correct boundary interaction, or that there
is no good conformal limit to the boundary Toda potential
(\ie the vertex operators in the boundary potential do not
correspond to operators in the conformal boundary spectrum).

Finally, two thoughts from extreme ends of the subject. First, it is
intriguing that the boundary action of the $N=2$ supersymmetric model
depends explicitly upon the bulk superpotential, and {\it not} just
on some derivatives of it. That is, the bosonic boundary potential is
$|W|^2$. In the discussion of the applications of singularity theory
to $N=2$ supersymmetric models, a minor frustration was that it was
hard to see a direct action of the monodromy group of the singularity
on the physical states of the theory. This is basically because the
physical states cared primarily about the critical points of $W$, and
not about the values of $W$, whereas the most beautiful theorems of
singularity theory relate to the latter. Since the infra-red limit of
the boundary theory is clearly concerned with the values of $W$, it
is conceivable that it might provide a clean physical realization of
the action of the monodromy group of the singularity.

The other thought is that it will be very interesting to see whether
the results presented here can be used in the study of monopoles or
cosmic strings in $(3+1)$ dimensions. As mentioned in the
introduction, treating a supersymmetric monopole as an impurity
problem will result in a potentially supersymmetric
$(1+1)$-dimensional boundary theory. Apart from the issue of quantum
integrability, it would also be very interesting to understand, in
the $(3+1)$ dimensional context, the interplay between boundary
degrees of freedom (excitations of the monopole), the amount of
supersymmetry, and the consequent Bogolmolnyi bounds.

\vskip 1.0 cm

\leftline{\bf Acknowledgements:}
\medskip
I would like to thank P.~Dorey, J.~McCarthy, J.-B.~Zuber and
particularly P.~Fendley and H.~Saleur for valuable discussions on
this work. I am also grateful to the the University of Adelaide, the
Service de Physique Th\'eorique at Saclay and to the High Energy
Physics Laboratory of the University of Paris VI in Jussieu for
hospitality while substantial parts of this work were done.

\vskip 1.0 cm

\leftline{\bf Dedication:}
\medskip
As this work was approaching completion I learned that Claude
Itzykson had died.  Apart from being a remarkable physicist, he
was also a good friend and great colleague.
I shall miss his conversation and
lectures, and regret that the exclamation ``excellent!'' will no
longer be used as an exhortation to rapidly terminate a boring
discussion.  I would like to dedicate this work,
such as it is, to Claude's memory.

\vfill
\eject
\listrefs

\vfill
\eject
\end